%% file: gevp.tex
\documentclass[11pt]{article}
\usepackage{amsmath,amsfonts,sint,epsfig,cite,graphics}
\usepackage{macros_static}
\usepackage{wrapfig}

\begin{document}

\input title.tex
\input s1.tex

\input s2.tex

\input s3.tex

\input s4.tex

\vskip2ex\noindent
{\bf Acknowledgements.}
The authors are grateful to Ferenc Niedermayer and Peter Weisz
for sharing their private note on the GEVP with us.
We thank Nicolas Garron for useful discussions
and an intense collaboration in the HQET project.
This work is supported
by the Deutsche Forschungsgemeinschaft
in the SFB/TR~09,
and by the European community
through EU Contract No.~MRTN-CT-2006-035482, ``FLAVIAnet''.
T.M. also thanks the A. von Humboldt Foundation for support.

\begin{appendix}

\input a1.tex

\end{appendix}


\providecommand{\href}[2]{#2}\begingroup\raggedright\endgroup

\end{document}

%% file: title.tex
\begin{titlepage}

\begin{flushright}
\vskip 0.7cm
DESY 09-014 \\
SFB/CPP-09-10\\
MKPH-T-09-01\\
LPT-Orsay/09-05\\
\end{flushright}

\vskip 0.35cm
\begin{center}
{\Large\bf 
On the 
generalized eigenvalue method for energies and matrix elements
in lattice field theory
}
\end{center}
\vskip 0.35cm
\vbox{
\centerline{
\epsfxsize=2.8 true cm
\epsfbox{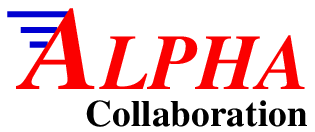}}
}
\vskip 0.1cm
\begin{center}
{
Benoit Blossier$^{\scriptscriptstyle a,b}$,
Michele Della Morte$^{\scriptscriptstyle c,d}$,
Georg von Hippel$^{\scriptscriptstyle a}$,
Tereza Mendes$^{\scriptscriptstyle a,e}$ and
Rainer Sommer$^{\scriptscriptstyle a}$ 
}
\vskip 0.5cm
{
\vskip 2.0ex
$^{\scriptstyle a}$
DESY,
Platanenallee 6, 15738 Zeuthen, Germany
\vskip 2.0ex
$^{\scriptstyle b}$
Laboratoire de Physique Th\'eorique,
B\^atiment 210, Universit\'e Paris XI,
F-91405~Orsay~Cedex, France
\vskip 2.0ex
$^{\scriptstyle c}$
CERN,
Physics Department, TH Division, CH-1211~Geneva~23, Switzerland
\vskip 2.0ex
$^{\scriptstyle d}$
Institut f{\"u}r Kernphysik, University of Mainz,
D-55099~Mainz, Germany
\vskip 2.0ex
$^{\scriptstyle e}$
IFSC, University of S\~ao Paulo, 
C.P.~369, CEP~13560-970, S\~ao~Carlos~SP, Brazil
}
\vskip 0.775cm
{\bf Abstract}
\vskip 0.1ex
\end{center}
We discuss the
generalized eigenvalue problem for computing energies
and matrix elements in lattice gauge theory, 
including effective theories such as HQET.
It is analyzed how the extracted effective energies
and matrix elements converge when the time separations
are made large. This suggests a particularly efficient
application of the method for which we can prove
that corrections vanish asymptotically as $\exp(-(E_{N+1}-E_n)\,t)$.
The gap $E_{N+1}-E_n$ can be made large by increasing
the number $N$ of interpolating fields in the correlation
matrix. \\
We also show how excited state matrix
elements can be extracted such that contaminations from all
other states disappear exponentially in time.\\
As a demonstration we present numerical results
for the extraction of ground state and excited B-meson masses
and decay constants
in static approximation and to order $1/m_{\rm b}$ in HQET.

\enlargethispage{2ex}
\vskip 2.0ex
\noindent{\it Key words:}
Lattice QCD; Heavy Quark Effective Theory; Hadronic Energies;
Hadronic Matrix Elements; Generalized Eigenvalue Problem
\vskip 2.0ex
\noindent{\it PACS:}
02.10.Ud; 
02.60.Dc; 
12.38.Gc; 
12.39.Hg; 
14.40.Nd  

\end{titlepage}

%% file: s1.tex
\newcommand{\corren}{\varepsilon_{n}}
\newcommand{\corrpn}{\pi_{nn'}}
\newcommand{\eneff}{E_n^\mrm{eff}}
\newcommand{\qeff}{{\cal \hat Q}_n^\mrm{eff}}
\newcommand{\qeffp}{{\cal \hat Q}_{n'}^\mrm{eff}}
\newcommand{\qeffg}{{\cal \hat Q}_1^\mrm{eff}}
\newcommand{\qefff}{{\cal Q}_n^\mrm{eff}}
\newcommand{\qefffp}{{\cal Q}_{n'}^\mrm{eff}}
\newcommand{\qn}{{\cal \hat A}_n}
\newcommand{\qnt}{{\cal \tilde A}_n}
\newcommand{\aeff}{{\cal \hat A}_n^\mrm{eff}}

\section{Introduction}
In the early days of lattice gauge theories, K. Wilson suggested to use
a variational technique to compute energy levels
in lattice gauge theory \cite{variational:Wilson}. The general idea
was quickly picked up and applied to the glueball
spectrum\cite{gevp:berg,gevp:michael} and to the static quark
potential(s)\cite{gevp:pot}. 

The usual choice of a 
variational basis starts from some interpolating fields
$O_i(x_0)$ already projected to a definite momentum and other 
quantum numbers such as parity. (Examples are given in the section
on numerical results.) The associated Hilbert space operators $\hat O_i$ 
define states
\be
  |\tilde\phi_i\rangle = \hat O_i|0\rangle\,
  \quad \mbox{and} \quad
  |\phi_i\rangle =\rme^{-t_0\, \hat H/2} |\tilde\phi_i\rangle \,,  
\ee
where $t_0$ is some time parameter and $\hat H$ the Hamilton operator.
A variational principle is formulated by ($t>t_0$)
\be
\lambda_1(t,t_0) = \Max{\{\alpha_i\}}\,
  {\langle \phi |\rme^{-(t-t_0)\hat H} | \phi \rangle \over
\langle \phi | \phi \rangle}\,, \quad
|\phi\rangle = \sum_{i=1}^N \alpha_i |\phi_i\rangle\,.
\ee
Clearly $\lambda_1(t,t_0)$ yields a variational estimate
of the lowest energy, $E_1$, with quantum numbers of $O_i$ via 
$\lambda_1(t,t_0)\approx \rme^{-E_1 (t-t_0)}$.
At the same time $\lambda_1(t,t_0)$ is the largest eigenvalue
of a generalized eigenvalue problem (GEVP), which we define below.

The GEVP is applicable beyond the computation of the ground state energy
and it has been widely used in lattice field theories
\cite{DeGrand:2006zz}. 
Recent examples are found in
\cite{Cais:2008cx,Bulava:2008pv,Prelovsek:2008rf,Burch:2008qx,Gattringer:2008vj,Bali:2008an,Gattringer:2008be,Ehmann:2007hj,Danzer:2007bx,Basak:2007kj,Dudek:2007wv,Foley:2007ui,Burch:2007fj}.
Corrections to the true energy levels
decrease exponentially for large time $t$,
when the energies are extracted from the 
generalized eigenvalues as described below \cite{phaseshifts:LW}. 

Apart from the paper of L\"uscher and Wolff~\cite{phaseshifts:LW}, 
statements about corrections due to other  
energy levels seem to be absent in the literature. We here add such statements 
and suggest a special use of the GEVP, which we will show 
to be more efficient under certain conditions. We also treat the case of an
effective theory and show numerical results for HQET. 
Furthermore, we show that the eigenvectors of the GEVP allow
to construct operators (see \eq{e:qeff}) which applied to the vacuum yield 
states converging exponentially to eigenstates of the 
Hamiltonian. 

With a proper condition for $t,t_0$ we prove that 
the convergence rate is $\sim \rme^{-(E_{N+1}-E_n) t} $ for the energies
and $\sim \rme^{-(E_{N+1}-E_n) t_0} $ for matrix elements of
the constructed states. The occurrence of the large gap $E_{N+1}-E_n$
is the main point of this paper.
An earlier presentation of our work on this topic can be found
in \cite{Blossier:2008tx}.

\begin{figure}[t]
\begin{center}
\includegraphics[height=0.9\textwidth,angle=270]{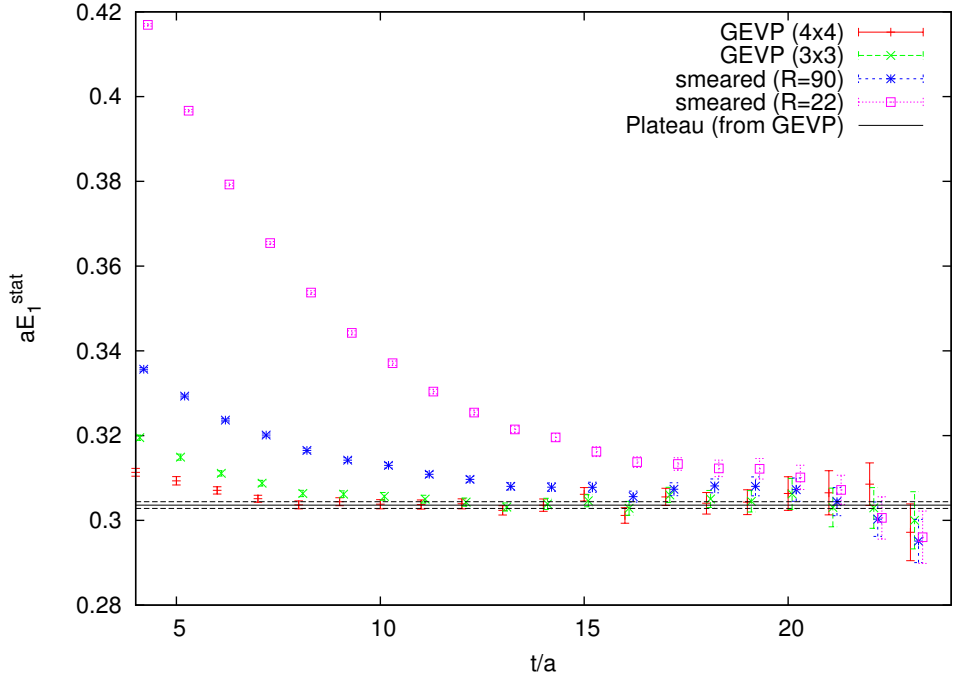}
\end{center}
\caption{Plot of $aE_1^{\rm stat}(t,t_0=4a)$ against $t/a$
         as obtained from the GEVP for a $3\times 3$ and a $4\times 4$
         system, and from the effective mass plot of two correlators
         with different amounts of Gaussian smearing (22 and 90
         iterations, respectively). The improved convergence of the
         GEVP solutions can be seen clearly.
}
\label{fig:compare}
\end{figure}

An indication of the effectiveness of the GEVP in suppressing
contaminating contributions from excited states can be seen in
figure \ref{fig:compare}, where we plot the effective energy of
the ground state of a static-light pseudoscalar meson as obtained
from different levels of Gaussian smearing and from the GEVP; for
details, we refer to section \ref{s:stat}.

\section{The Generalized Eigenvalue Problem \label{s:gevp}}
\subsection{The basic idea}

We start from a matrix of Euclidean space correlation functions 
\bes
  \label{e:cij}
  C_{ij}(t) &=& \langle O_i(t) O_j^*(0) \rangle =
  \sum_{n=1}^\infty \rme^{-E_n t} \psi_{ni}\psi_{nj}^*\,,\quad 
  i,j = 1,\ldots,N  \\ && \quad \psi_{ni} \equiv (\psi_n)_i = 
  \langle 0|\hat O_i|n\rangle
  \quad
  E_n < E_{n+1} \,.
 \nonumber 
\ees 
Note that we have assumed non-degenerate energy levels.
Space-time could be continuous, but in practice applications
will be for discretized field theories. In that case, we 
assume that either the theory has a hermitian, positive transfer 
matrix as for the standard Wilson gauge theory \cite{Luscher:TM,Creutz:tm},
or we consider $t,t_0$ large enough and the lattice spacing, $a$, 
small enough 
such that the correlation functions are well represented
by a spectral representation and complex energy contributions
are irrelevant \cite{impr:tm,silvia:universality}.
States $|n\rangle$ with
$\langle m|n\rangle =\delta_{mn}$ are eigenstates of the Hamiltonian
(logarithm of the transfer matrix)
and all energies, $E_n$, have the vacuum energy subtracted.
$O_j(t)$ are fields on a timeslice $t$ that correspond
to Hilbert space operators $\hat O_j$ whose quantum numbers 
such as parity and flavor numbers are
then also carried by the states $|n\rangle\,, n>0$. In the above formula,
contributions due to a finite time extent 
of space-time are neglected, so the time extent $T$ has to be 
large.~\footnote{See \cite{gevp:francesco} for a discussion of finite $T$
when the $O_i$ carry vacuum quantum numbers.} 

Before discussing the GEVP, let us comment briefly on a wider scope of 
applications.
Besides the energy levels $E_n$ one may want to determine 
matrix elements $\langle 0|\hat P|n\rangle$ or $\langle m|\hat P|n\rangle$
of operators $\hat P$ that may or may not be in the set of
operators $\{ \hat O_i \}$. We will see that the
GEVP provides a systematic approach to this problem as well.

The GEVP is defined by, 
\bes \label{e:gevp}
  C(t)\, v_n(t,t_0) = \lambda_n(t,t_0)\, C(t_0)\,v_n(t,t_0) \,, 
  \quad n=1,\ldots,N\,,\quad t>t_0\,.  
\ees
L\"uscher and Wolff showed that 
one can use it to determine systematically also excited
states, namely
\cite{phaseshifts:LW} 
\bes
  E_n &=& \lim_{t\to\infty} E_n^\mrm{eff}(t,t_0)\,,\\
  \label{e:eneff1}
  E_n^\mrm{eff}(t,t_0) &=& -\partial_t \log\lambda_n(t,t_0) \equiv 
    -{1\over a} \, [\log{\lambda_n(t+a,t_0)}-\log{\lambda_n(t,t_0)}]\,.
\ees

For a while (until \sect{s:gen}) we now assume a simplified case, which helps
to understand the usefulness of $E_n^\mrm{eff}$ and
$v_n$. The simplification is that only $N$ states contribute,
\bes
  C_{ij}(t)= C_{ij}^{(0)}(t) = \sum_{n=1}^N \rme^{-E_n t} \psi_{ni}\psi_{nj}^*
                  \,. \label{e:C0}
\ees
We introduce the dual (time-independent) vectors $u_n$, defined by
\bes
  (u_n,\psi_m) = \delta_{mn}\,,\quad m,n\leq N\,
 \label{e:dual}
\ees
with $(u_n,\psi_m)\equiv \sum_{i=1}^N (u_n^*)_i\,\psi_{mi}$. Inserting into
\eq{e:C0} gives
\bes
  C^{(0)}(t) u_n = \rme^{-E_n t} \psi_n \,, \label{e:vn0}\quad
  C^{(0)}(t)\, u_n = \lambda_n^{(0)}(t,t_0)\, C^{(0)}(t_0)\,u_n\,.
                                \label{eq:gevp0}
\ees
So the GEVP is solved by
\bes
  \lambda_n^{(0)}(t,t_0) = \rme^{-E_n (t-t_0)}\,, \quad
  v_n(t,t_0) \propto u_n
\label{eq:lambda0}
\ees
and there is an orthogonality for all $t$ of the form
\bes
  (u_m,C^{(0)}(t)\,u_n) = \delta_{mn}\,\rho_n(t)\,,
  \quad \rho_n(t) = \rme^{-E_n t}\,.
  \label{e:norm}
\ees
These equations mean that the hermitean conjugates of the operators
\be
  \qn = \sum_{i=1}^N (u_n^*)_i \hat O_i \equiv (u_n,\hat O)\,,
\ee
create the eigenstates of the Hamilton operator, 
\be
  |n\rangle = \qn^\dagger |0 \rangle \,,\quad \hat H|n \rangle =E_n\,|n \rangle\,.
\ee
Consequently matrix elements are easily obtained. For example
\be
    \langle 0|\hat P|n\rangle =  \langle 0| \hat P \qn^\dagger |0\rangle\,, 
\ee
translates into correlation functions of $P,O_i$. 

\subsection{General statements about energies and matrix elements \label{s:gen}}

Let us now come back to the general case \eq{e:cij}. The idea is to solve
the GEVP, \eq{e:gevp}, ``at large time'' where the contribution of 
states $n>N$ is
small.
We then consider the effective energies, \eq{e:eneff1},
and effective operators\footnote{For simplicity we discuss a 
special case of a more general class of operators depending 
on several times, which then all have to be taken large. An example is
\bes
    \label{e:qeffp}
    \qeff &=& R_n \,(\hat O\,,\,v_n(t_1,t_0)\,) \,,\\
    R_n &=&
               \left(v_n(t_1,t_0)\,,\, C(t_2)\,v_n(t_1,t_0)\right)^{-1/2}
               {\lambda_n(t_0+t_2/2,t_0) \over \lambda_n(t_0+t_2,t_0)}\,.
\ees
}
\bes
    \aeff(t,t_0) &=& \rme^{-\hat H t} \qeff(t,t_0) \,,\\
    \label{e:qeff}
    \qeff(t,t_0) &=& R_n \,(\hat O\,,\,v_n(t,t_0)\,) \,,\\
    R_n &=&
               \left(v_n(t,t_0)\,,\, C(t)\,v_n(t,t_0)\right)^{-1/2}
               {\lambda_n(t_0+t/2,t_0) \over \lambda_n(t_0+t,t_0)}\,.
\ees
Corrections to the large time asymptotics are parameterized by
\bes
  \label{e:eneff}
  E_n^\mrm{eff}(t,t_0) &=&  E_n + \corren(t,t_0)\,, \\
   \rme^{-\hat H t}(\qeff(t,t_0))^\dagger|0\rangle &=&
       |n\rangle + \sum_{n'=1}^\infty  \corrpn(t,t_0) |n'\rangle
       \,.
\ees
The terms $\corren,\corrpn$ will disappear exponentially at large times.
Note that in the literature
the energy levels are sometimes not extracted like that and
rather the standard effective masses of
correlators made from $\qnt =(O,v_n(t_1,t_0))$ are used 
with fixed $t_1,t_0$
and the question
of the size of the corrections is left open.
However, the form \eq{e:eneff} has a theoretical advantage as it
was shown in
 \cite{phaseshifts:LW} that (at fixed $t_0$)
\bes
  \corren(t,t_0) = \rmO(\rme^{-\Delta E_n\, t})\,,\quad \Delta E_n =
  \min_{m\neq n}\, |E_m-E_n|\,.
  \label{e:LWformula}
\ees
This is non-trivial as it allows to obtain the excited levels
with corrections which vanish in the limit of large $t$, keeping $t_0$ fixed.
However, it appears from this formula that the corrections can be large
when there is an energy level close to the desired one. This is the case
in interesting phenomena such as string breaking 
\cite{pot:Higgs1,pot:Higgs2,Philipsen:1998de,Bali:2005fu,Pepe:2009in},
where in numerical  applications the
corrections appeared to be very small despite the formula 
above.
Also in static-light systems the gaps
are typically only around $\Delta E_n \approx 400\,\MeV$,
and in full QCD with light quarks, a small gap 
 $\Delta E_n \approx 2 m_\pi$ appears.

Our contribution to the issue is a more complete discussion of the 
correction $\corren$ to $E_n$ as well as the definition of the
effective operator, \eq{e:qeff}, 
and a discussion of the corrections $\corrpn$ to its
matrix elements. It turns out that in all this, a very useful case is to consider
the situation 
\be
  t_0 \geq t/2\,,
\ee
e.g.\ with $t-t_0 = \mbox{const.}$ or $2\geq t/t_0=\mbox{const.}$ 
and take $t_0$ (in practice moderately) large. 
In this region of $t,t_0$ the second order perturbations 
in higher states $n>N$ are not larger than the first order ones.
We can then show that 
\bes
  \label{e:corren}
  \corren(t,t_0) &=& \rmO(\rme^{-\Delta E_{N+1,n}\, t}) \,,\quad 
                   \Delta E_{m,n} = E_m-E_n \,, \\
  \label{e:corrpn}
  \corrpn(t,t_0) &=& \rmO(\rme^{-\Delta E_{N+1,n}\, t_0})\,,\quad \mbox{at
    fixed } t-t_0 \\
  \label{e:corrp1}
  \pi_{1n}(t,t_0) &=& \rmO(\rme^{-\Delta E_{N+1,1}\, t_0} 
  \rme^{-\Delta E_{2,1}\,(t- t_0)}) \,+\, \rmO(\rme^{-\Delta E_{N+1,1}\, t}) 
   \,.
\ees
The large gaps $\Delta E_{N+1,n}$ can solve the problem of close-by levels
for example in the string-breaking situation, but also speed
up the general convergence very much. For example in static-light systems
$\Delta E_{6,1} \approx 2\,\GeV$ translates into a gain of a factor of 5 in time
separation.  

\Eq{e:corrpn} means that matrix elements of (time-)local operators $\hat P$
and the associated field $P$
can be computed via
\bes
  \langle0| \qeff  \rme^{-\hat H t}\hat P  \rme^{-\hat H t}(\qeffp)^\dagger |0\rangle =
  \langle \qefff(2t) P(t) (\qefffp(0))^* \rangle
  = \langle n | \hat P | n'\rangle +  \rmO(\rme^{-\Delta E_{N+1,n}\, t_0})
  \nonumber \\
\ees
and similarly for $ \langle n | \hat P | 0\rangle$.

We now turn to an outline of the proof of these
statements, delegating the technical part to an
appendix.

%% file: s2.tex
\section{Perturbation theory \label{s:pt}}
We start from the solutions above
for $C=C^{(0)}$ and treat the higher states as perturbations. 
This perturbative evaluation was 
already set up by F. Niedermayer and P. Weisz a while 
ago \cite{notes:FerencPeter} but never published. We noted 
the r\^ole of $ t_0 \geq t/2$, the form of the corrections
to the effective operators defined above and could show
that these relations hold to all orders in the
perturbative expansion.    

We want to obtain $\lambda_n$ and ${v}_n$ in a perturbation
theory in $\eps$, where
\be 
  \label{e:pgevp}
  A{v}_n = \lambda_n B {v}_n\, ,\quad A=A^{(0)}+\eps A^{(1)} \, ,\quad
  B=B^{(0)}+\eps B^{(1)} \,.
\ee
All these $N\times N$ matrices are assumed to be hermitian, 
$B$ and $B^{(0)}$ are assumed to be positive.
We will set
\bea
  \label{e:A}
  A^{(0)}&=& C^{(0)}(t)\,,\quad \eps A^{(1)}= C^{(1)}(t) \,,\\ 
  \label{e:B}
  B^{(0)}&=& C^{(0)}(t_0)\,,\quad \eps B^{(1)}= C^{(1)}(t_0) \,
\ees
in the end with
\be
  \label{e:cij1}
  C_{ij}^{(1)}(t) \;=\;
  \sum_{n=N+1}^\infty \rme^{-E_n t} \psi_{ni}\psi_{nj}^*\,,
\ee
which is hermitian and positive under the assumptions listed in the 
previous section.
A possible choice for $\eps$ is $\eps=\rme^{-(E_{N+1}-E_{N}) t_0}$ 
which becomes arbitrarily small for large $t_0$.
The solutions of the lowest-order equation
\be
  A^{(0)}{v}_n^{(0)} \;=\; \lambda_n^{(0)} B^{(0)}\, {v}_n^{(0)}\,,
\ee
satisfy an orthogonality relation 
\be
\,  ({v}_n^{(0)},B^{(0)}{v}_m^{(0)}) = \rho_n \, \delta_{nm}\,
  \label{eq:rho}
\ee
 as in \eq{e:norm} above. Non-degenerate eigenvalues 
$\lambda_n^{(0)} > \lambda_{n+1}^{(0)}$
are assumed.
Expanding 
\bes
  \lambda_n \;=\; \lambda_n^{(0)}+\eps \lambda_n^{(1)} +\eps^2 \lambda_n^{(2)}\,\ldots 
  \,,\quad
  {v}_n \;=\; {v}_n^{(0)}+\eps {v}_n^{(1)} +\eps^2 {v}_n^{(2)}\,\ldots 
\ees
with the orthogonality condition
\bes
\,  (v_n^{(k)}, B^{(0)} v_n^{(0)}) = 0\,,\quad  k=1,2,\dots\,,
\ees
we get for the first two orders
\bes
A^{(0)} {v}_n^{(1)} + A^{(1)} {v}_n^{(0)}  &=&
\lambda_n^{(0)}\,\left[B^{(0)} {v}_n^{(1)} + B^{(1)} {v}_n^{(0)}\right]
+\lambda_n^{(1)}\,B^{(0)} {v}_n^{(0)} \,, \label{eq:1stord} \\[1mm]
A^{(0)} {v}_n^{(2)} + A^{(1)} {v}_n^{(1)}  &=&
\lambda_n^{(0)}\,\left[B^{(0)} {v}_n^{(2)} + B^{(1)} {v}_n^{(1)}\right]
+\lambda_n^{(1)}\,\left[B^{(0)} {v}_n^{(1)} + B^{(1)} {v}_n^{(0)}\right]
\nonumber \\ &&
+\lambda_n^{(2)}\,B^{(0)} {v}_n^{(0)} \,. \quad \label{eq:2ndord} \quad
\ees

With the orthogonality, \eq{eq:rho},
of the lowest order vectors, one obtains 
just like in 
ordinary quantum mechanics perturbation theory the solutions for eigenvalues and
the eigenvectors
\bes
  \label{e:lam1}
  \lambda_n^{(1)} &=&  \rho_n^{-1}\,
                  \left({v}_n^{(0)},\Delta_n {v}_n^{(0)}\right)\,,
                  \quad \Delta_n \,\equiv\, A^{(1)} - \lambda_n^{(0)} B^{(1)}\, \\
{v}_n^{(1)}  &=& \sum_{m\neq n} \alpha_{nm}^{(1)}\,{v}_m^{(0)} \,, 
\label{e:alphanm} \quad
\alpha_{nm}^{(1)}  =  \rho_m^{-1}
 { \left({v}_m^{(0)},\Delta_n {v}_n^{(0)}\right) \over
                   \lambda_n^{(0)} - \lambda_m^{(0)} }
\\
  \label{e:lam2}
  \lambda_n^{(2)} &=& \sum_{m\neq n} \rho_n^{-1} \rho_m^{-1}  
                 { \left|\left({v}_m^{(0)},\Delta_n {v}_n^{(0)}\right)\right|^2 \over
                   \lambda_n^{(0)} - \lambda_m^{(0)} } 
                 - \rho_n^{-2}\left({v}_n^{(0)},\Delta_n {v}_n^{(0)}\right)
                   \left({v}_n^{(0)},B^{(1)} {v}_n^{(0)}\right) \,.
\ees
A recursion formula for the higher-order
coefficients is given in \app{a:recursion}.

We note that for our case of interest, $t_0$ can be
chosen large enough such that the perturbation theory is
absolutely convergent. It therefore suffices to discuss the large $t_0,t$ 
asymptotics order by order in the expansion in order
to show the detailed form of corrections, in particular that they are 
controlled by the large gap 
$\Delta E_{N+1,n}$. Convergence of the expansion 
is guaranteed for the following reason. As in quantum mechanics, 
the perturbation theory can be
set up with the help of the resolvent depending on a complex
variable $z$. The positions of its poles yield the (here generalized) 
eigenvalues. The resolvent for the GEVP is 
\be
  G(z) \;=\; (zB-A)^{-1} \;=\; G_0(z) \sum_{k=0}^\infty\, \eps^k\, 
  [(A^{(1)} - zB^{(1)})\,G_0(z)]^k \,, \label{e:resolvent}
\ee
in terms of the unperturbed resolvent $G_0(z)= (zB^{(0)} -A^{(0)})^{-1}$. 
The series expansion \eq{e:resolvent} converges for large $t_0$ 
(and $t>t_0$) since 
$||\eps (zB^{(1)} -A^{(1)})|| < ||G_0^{-1}(z)||$ holds at 
sufficiently large $t_0$ (note that the resolvent is needed only 
away from the zeros of $G_0^{-1}(z)$).

\subsection{Application to the perturbations $C^{(1)}$ }
Now we insert our specific problem \eq{e:A}, \eq{e:B}. 
With a representation (for $m>n$)
\bes
    (\lambda_n^{(0)} - \lambda_m^{(0)})^{-1} &=&
    (\lambda_n^{(0)})^{-1}(1-\rme^{-(E_m-E_n)(t-t_0)})^{-1}
  \nonumber \\
    &=&  (\lambda_n^{(0)})^{-1} \sum_{l=0}^\infty \rme^{-l(E_m-E_n)(t-t_0)}\,,
\ees
and some algebra spelled out in \app{a:algebra} 
one can show that eqs.(\ref{e:corren} -- \ref{e:corrp1})
hold at any order in $\eps$.

\input tereza.tex

\subsection{Effective theory to first order}
\def\first{\mrm{1/m}}
\def\stat{\mrm{stat}}
In an effective theory such as HQET, all correlation functions
\bes
  C_{ij}(t) &=& C_{ij}^{\stat}(t) \,+\, 
  \omega \,C_{ij}^{\first}(t) \,+\, \rmO(\omega^2)
\ees
are computed in an expansion in a small parameter, $\omega$,
which we here consider to first order only. The notation is taken
from HQET where $\omega \propto \minv$. It also helps to avoid 
a confusion with the perturbation theory in terms of $C^{(1)}$ 
treated before.

\subsubsection{Energies}

We start our analysis from the GEVP in the full theory, \eq{e:gevp},
and use the form of the correction terms of the effective energies 
($t\leq2t_0$)  
\bes
    E_n^{\rm eff}(t,t_0) \;=\; a^{-1}\,\log {\lambda_n(t,t_0) \over \lambda_n(t+a,t_0)} \;=\; 
    E_n + \rmO(\rme^{-\Delta E_{N+1,n}\, t}),
\ees
see the discussion above.
Expanding this equation in $\omega$, we arrive at
\bes 
    E_n^{\rm eff}(t,t_0) &=& E_n^{\rm eff,\stat}(t,t_0) 
     +\omega E_n^{\rm eff,\first}(t,t_0) +\rmO(\omega^2)\\
    E_n^{\rm eff,\stat}(t,t_0) &=& a^{-1}\,\log {\lambda_n^\stat(t,t_0) \over 
                         \lambda_n^\stat(t+a,t_0)} 
   \label{e:lamstat}
 \\[1mm]
     E_n^{\rm eff,\first}(t,t_0) &=& {\lambda_n^\first(t,t_0) \over 
                         \lambda_n^\stat(t,t_0)} \,-\,
      {\lambda_n^\first(t+a,t_0) \over 
                         \lambda_n^\stat(t+a,t_0)} 
   \label{e:lamfirst}
\ees
with the behavior at large time,
\bes
    E_n^{\rm eff,\stat}(t,t_0) &=&
    E_n^\stat \,+\, \beta_n^\stat\,\rme^{-\Delta E_{N+1,n}^\stat\, t}+\ldots\,,
   \label{e:lamstatfit}
  \\
    E_n^{\rm eff,\first}(t,t_0) &=& 
    E_n^\first \,+\, [\,\beta_n^\first 
               \,-\, \beta_n^\stat\,t\,\Delta E_{N+1,n}^\first\,] 
                     \rme^{-\Delta E_{N+1,n}^\stat\, t}+\ldots\, .
   \label{e:lamfirstfit}
\ees
Following the beginning of \sect{s:pt} we now take the lowest order
correlator to define the unperturbed GEVP,
\bes
    C^\stat(t) \,v_n^\stat(t,t_0) \;=\; \lambda_n^\stat(t,t_0)\, C^\stat(t_0)
             \,v_n^\stat(t,t_0) \,, 
\ees
whose eigenvectors with normalization
$
          (v_m^\stat(t,t_0)\,,\,C^\stat(t_0)\,v_n^\stat(t,t_0)) =\delta_{mn}\,
$
are then needed in the formula 
\bes
  {\lambda_n^\first(t,t_0)\over \lambda_n^\stat(t,t_0)}
  &=& \left(v_n^\stat(t,t_0)\,,\,
             [[\lambda_n^\stat(t,t_0)]^{-1}\,C^{\first}(t)- C^{\first}(t_0)] 
             v_n^\stat(t,t_0)\right)
\ees
for the first-order corrections in $\omega$. We note that the 
finite time corrections
in \eq{e:lamfirst} comprise of both a term $\rme^{-\Delta E_{N+1,n}^\stat\, t}$ and
one with an extra factor of $t$, however only their prefactors are new,
once $\Delta E_{N+1,n}^\stat$ is known 
from an analysis of the lowest order eigenvalues. 
Due to the large gap $\Delta E_{N+1,n}^\stat$, the corrections 
disappear quickly with time.

\subsubsection{Matrix elements}

The discussion of matrix elements starts from the operators
$\qeff$, \eq{e:qeff}. Their expansion
\be
 \qeff = R_n^\stat \,(v_n^\stat(t,t_0)\,,\,\hat O) + \omega\,\left[
         R_n^\first \,(v_n^\stat(t,t_0)\,,\,\hat O) \,+\,
         R_n^\stat \,(v_n^\first(t,t_0)\,,\,\hat O)  \right]
\ee
is given in terms of 
\bes 
    R_n^\stat &=&
               \left(v_n^\stat(t,t_0)\,,\, C^\stat(t)\,v_n^\stat(t,t_0)
               \right)^{-1/2}
               {\lambda_n^\stat(t_0+t/2,t_0) \over
                 \lambda_n^\stat(t_0+t,t_0)}\,,
    \label{e:rstat}
   \\
   {R_n^\first \over R_n^\stat}&=&
        - {1\over 2}
       {\left(v_n^\stat(t,t_0),\, C^\first(t)\,v_n^\stat(t,t_0)\right)
     \over \left(v_n^\stat(t,t_0),\, C^\stat(t)\,v_n^\stat(t,t_0)\right)}
    \, \nonumber \\ &&
    \;+\;
          {\lambda_n^\first(t_0+t/2,t_0) \over
                         \lambda_n^\stat(t_0+t/2,t_0)}
              -{\lambda_n^\first(t_0+t,t_0) \over
                         \lambda_n^\stat(t_0+t,t_0)}   
      \label{e:rfirst}
\ees
and the first order perturbation to $v_n$,
\be
  v_n^\first = \sum_{k\ne n = 1}^N v_k^\stat
      { \left({v}_k^\stat, [C^{\first}(t)-\lambda_n^\stat(t,t_0)C^{\first}(t_0)]\,
         {v}_n^\stat\right) \over
               \lambda_n^\stat(t,t_0)-\lambda_k^\stat(t,t_0) }\,.
\ee
A simple matrix element is then 
\bes 
   p_n^{\rm eff}(t,t_0) = \langle0|\qeff \rme^{-\hat H t}  \hat P |0\rangle
   &=&  R_n\, (v_n(t,t_0),C_\mrm{P}) \,,
   \quad (C_\mrm{P})_j = \langle  O_j(0) P(t) \rangle \,.
\ees
For example when $P$ denotes the time component of an axial current,
it yields the decay constant of the associated ground state meson via,
\be
   p_1 = \langle 0 | \hat P | 1 \rangle\,,
\ee
with the $\omega$-expansion
\bes
   p_n^{\rm eff}(t,t_0) &=&  p_n^{\rm eff,stat}(t,t_0)\,
  (1 + \omega\,  p_n^{\rm eff,\first}(t,t_0)
                    + \rmO(\omega^2)) \\
   p_n^{\rm eff,\first}(t,t_0) &=& {R_n^\first \over R_n^\stat}\,+\,
   {(v_n^\stat(t,t_0), C_\mrm{P}^\first(t)) \over (v_n^\stat(t,t_0),C_\mrm{P}^\stat(t))} \,+\, 
   {(v_n^\first(t,t_0),C_\mrm{P}^\stat(t)) \over (v_n^\stat(t,t_0),C_\mrm{P}^\stat(t))}\,.
   \label{e:psifirst}
\ees
Specializing further, 
the large time asymptotics of the effective ground state matrix element
has the following form
\bes
    p_1^{\rm eff}(t,t_0) &\sim&
    p_1 + \gamma \rme^{-\Delta E_{N+1,1}  t_0}\,, \\[1ex]
    \label{e:psistatfit}
    p_1^{\rm eff,\stat}(t,t_0) &\sim&
    p_1^\stat + \gamma^\stat\, \rme^{-\Delta E_{N+1,1}^\stat t_0}\,,
 \\[1mm]
    {p_1^{\rm eff,\first}(t,t_0)}
    &\sim&
    \label{e:psifirstfit}
    {p_1^\first}
    \,+\, \rme^{-\Delta E_{N+1,1}^\stat t_0}\, \left[
    {\gamma^\first}
    - {\gamma^\stat}{p_1^\first}
    - t_0\,\Delta E_{N+1,1}^\first {\gamma^\stat}\,
    \right]\,. \nonumber \\
\ees
We note that the correction term proportional to 
$t_0\, \rme^{-\Delta E_{N+1,1}^\stat t_0}$ is entirely fixed by a
first order analysis of the energies and the 
lowest order one of the matrix element. \\[3ex]

%% file: tereza.tex
In particular, we list below the explicit expressions up to second order in
$\epsilon$, keeping only leading terms in the sums. They demonstrate how the 
condition $t_0\geq t/2$ comes about.

For the effective energies we have
\bes
\varepsilon_n(t,t_0) &\sim& \rme^{-(E_{N+1}-E_{n})t}
   \left[1-  \rme^{-(E_{N+1}-E_{n}) a}\right] c_{n,n,N+1} \\[2mm]
   &&+\;  \sum_{m>n}  \nonumber
   \rme^{-(2 E_{N+1}-E_{n}-E_m) t_0}
   \rme^{-(E_{m}-E_{n}) (t-t_0)}
   \left[1-\rme^{-(E_{m}-E_{n}) a}\right] \left|c_{n,m,N+1}\right|^2 \\[2mm]
   &&-\;  \sum_{m<n}  \nonumber
   \rme^{-(2 E_{N+1}-E_{n}-E_m) t_0}
   \rme^{-(E_{n}-E_{m}) (t-t_0)}
   \left[1-\rme^{-(E_{n}-E_{m}) a}\right] \left|c_{n,m,N+1}\right|^2
\ees
where the first term (line) comes from order $\epsilon$ in the 
perturbative expansion
and the second and third terms come from order $\epsilon^2$. Also,
we have defined
\bes
        c_{n,m,l}= (u_n,\psi_{l}) (\psi_{l},u_m)\,.
\ees
We see that, due to cancellations of $t$-independent terms in the effective energy,
the first-order correction is independent of $t_0$ and exponentially 
suppressed in $t$ with the large 
energy gap $E_{N+1}-E_{n}$ as coefficient. (This correction is positive, while
the sign of the combined second-order one depends on the energy shifts 
for the states around $E_n$.)
At fixed $t_0$ and asymptotically large $t$ the second-order corrections 
dominate because they follow the slower decay \eq{e:LWformula}.
However, simple inspection of the above corrections shows that
the $t_0$-dependent prefactors supress the second order
correction sufficiently when $t_0\geq t/2$. Then the 
first-order one dominates, confirming
\eq{e:corren}.

For the amplitudes $\pi_{nn'}(t,t_0)$ we have main contributions 
at first order in $\epsilon$ for $n'\neq n$, $n'\leq N$. These
contributions from $n'>n$ and $n'<n$ are given respectively by
\bes
\pi_{nn'}(t,t_0) \;=\; - c_{n,n',N+1}\, 
      \rme^{-(E_{N+1}-E_{n}) t_0}\, \rme^{-(E_{n'}-E_{n})(t-t_0)}
\ees
and
\bes
\pi_{nn'}(t,t_0) \;=\; c_{n,n',N+1}\,
      \rme^{-(E_{N+1}-E_{n}) t_0} \,.
\ees
Note that in the $n'>n$ case the value $n'=n+1$ has a
leading exponential behavior, whereas for $n'<n$ all values of
$n'$ have the same exponential behavior.

%% file: s3.tex
\section{Application to static-light B$_\mrm{s}$-mesons \label{s:stat}}

We have carried out numerical tests of eqns.
(\ref{e:lamstat})-(\ref{e:psifirstfit})
in the context of static-light $B_s$ mesons in quenched HQET
with a HYP2 static quark
\cite{HYP,HYP:pot,DellaMorte:2005yc}
and a non-perturbatively $\rmO(a)$-improved
\cite{impr:SW,impr:pap3}
Wilson valence (strange)
quark\footnote{One may object that a quenched calculation does not
satisfy our basic premise, \eq{e:cij}. However the quark considered
here is rather heavy and experience has shown that quenching
represents a small modification of the full theory in such cases.
Clearly a demonstration in a unitary theory will be welcome.}.
Our lattices are $L^3\times T$ with $L\approx 1.5$ fm, $T=2L$ and
periodic boundary conditions in all directions.
For this demonstration we have chosen
$a=0.07$ fm ($\beta=6.2885$, $\kappa_s=0.1349798$)
and use an ensemble of 100 quenched configurations.
A more detailed analysis including multiple lattice spacings to
enable us to take the continuum limit is in progress.

Strange quark propagators are computed using a variant of
the Dublin method
\cite{alltoall:dublin}.
We use approximate instead of exact low modes and employ even-odd
preconditioning in order to reduce both the size of the eigenvalue
problem to be solved for the low modes and the number of noise sources
to be used for the stochastic estimator by a factor of 2; this leads
to a significant reduction in effort and is expected to also reduce
the amount of noise introduced by the stochastic estimation of the
short-distance contributions
\cite{Blossier:inprep}.

The interpolating fields are constructed using quark bilinears
\begin{eqnarray}
O_k(x) &=& \psibar_{\rm h}(x)\gamma_0\gamma_5\psi_{\rm l}^{(k)}(x) \\
O_k^*(x) &=& \psibar_{\rm l}^{(k)}(x)\gamma_0\gamma_5\psi_{\rm h}(x)
\end{eqnarray}
of identical Dirac structure, but with different levels of Gaussian
smearing
\cite{wavef:wupp1}
for the light quark fields
\begin{equation}
\psi_{\rm l}^{(k)}(x) = \left( 1+\kappa_G\Delta \right)^{R_k} \psi_{\rm l}(x)\,,
\end{equation}
where the gauge fields in the covariant Laplacian are first smeared
with 3 iterations of (spatial) APE smearing
\cite{smear:ape,Basak:2005gi}.
We use $R_k=22$, $45$, $67$, $90$, $135$, $180$, $225$ with
$\kappa_G=0.1$.
The local ($R_0=0$) operator is also included
in order to be able to compute the decay constant. Here and throughout,
$\psi_{\rm h}(x)$ denotes the static quark field.

For these operators, we compute the following correlators:
\bea\nonumber
C^{\rm{stat}}_{ij}(t)& = &{a^{13} \over L^6\,T} \sum_{\vec{x},\vec{y},t_x} 
\left< O_i(\vec{y},t+t_x)
O^*_j(\vec{x},t_x)\right>,\\
\nonumber
C^{\delta A}_i(t)& = &{a^{14} \over L^6\,T} \sum_{\vec{x},\vec{y},t_x} 
\left< O_i(\vec{y},t+t_x) O^*_{\delta A}(\vec{x},t_x)\right>,\\
\nonumber
C^{\rm{kin}}_{ij}(t)& = &{a^{21} \over L^9\,T} \sum_{\vec{x},\vec{y},\vec{z},
t_x,t_x\leq t_y\leq t_x+t}  
\left< O_i(\vec{z},t+t_x) O_{\rm kin}(\vec{y},t_y) O^*_j(\vec{x},t_x)\right>,\\
C^{\rm{spin}}_{ij}(t)& = &{a^{21} \over L^9\,T} \sum_{\vec{x},\vec{y},\vec{z},
t_x,t_x\leq t_y\le t_x+t} 
\left< O_i(\vec{z},t+t_x) O_{\rm spin}(\vec{y},t_y)O^*_j(\vec{x},t_x)\right>,
\eea
where 
\bea\nonumber
O_{\delta A}(\vec{x},t)&=&\psibar_{\rm h}(\vec{x},t) \gamma_0 \gamma_5 
[\vec{\gamma} \cdot \vec{D} \psi_{\rm l}^{(0)}](\vec{x},t),\\
\nonumber
O_{{\rm kin}}(\vec{x},t)&=&\psibar_{\rm h}(\vec{x},t) [\vec{D}^2 \psi_{\rm h}]
(\vec{x},t),\\
O_{{\rm spin}}(\vec{x},t)&=&\psibar_{\rm h}(\vec{x},t) 
[\vec{\sigma} \cdot \vec{B} \psi_{\rm h}](\vec{x},t).
\eea
The correlator $C^{\delta A}$ is introduced to ensure
the axial local current is correct up to both ${\cal O}(a)$
and ${\cal O}(1/m_{\rm b})$, whereas the last two correlators are
the ${\cal O}(1/m_{\rm b})$ terms from the HQET action.\\
The energy of a state $|n\rangle$ is given by
\be
E_n = E_n^{\rm{stat}} + \omega_{\rm kin} E_n^{\rm{kin}} + 
\omega_{\rm spin} E_n^{\rm{spin}} + \delta m
\ee
and its decay constant factorizes as
\be
f^{(n)}_B \sqrt{M^{(n)}_B/2}=Z_A^{\rm HQET} p^{\rm stat}_n (1+\omega_{\rm kin} p^{\rm kin}_n +
\omega_{\rm spin} p^{\rm spin}_n + c_A^{\rm HQET} p^{\delta A}_n)
\ee
where $Z_A^{\rm HQET}$, $\omega_{\rm kin}$, $\omega_{\rm spin}$,
$c_A^{\rm HQET}$ and $\delta m$ are (divergent) matching
constants between the QCD action and currents and their HQET counterpart.
A strategy to compute them has been presented in
\cite{DellaMorteCB, lat07:nicolas}
and we quote their approximate numerical values
\cite{Blossier:inprep}
at $\beta=6.2885$
($Z_A^{\rm HQET}\sim 1$, $\omega_{\rm kin}/a\sim0.4$,
$\omega_{\rm spin}/a\sim 0.7$, $c_A^{\rm HQET}/a\sim -0.6$)
only as an illustration of the expected size of their contributions.

The resulting $8\times 8$ correlator matrices $C(t)$ are symmetrized
and then truncated to $N\times N$ matrices $C^{(N\times N)}(t)$
by projecting with the $N$ eigenvectors belonging to the $N$ largest
eigenvalues of $C^{\rm stat}(t_i)$:
\begin{eqnarray}
C^{\rm stat}(t_i) b_n &=& \lambda_n b_n \\
C^{(N\times N)}_{nm}(t) &=& b_n^\dag C(t) b_m\,,\;n,m\leq N.
\end{eqnarray}
For $N$ not too large, this helps to avoid numerical instabilities in
the GEVP that could otherwise lead to large errors
\cite{gevp:bern}.
We use $t_i=2a$.

For each of the resulting $N\times N$ correlators, we solve the static GEVP
and compute the static and $\rmO(1/m_{\rm b})$ energies and matrix elements
as per eqns.
(\ref{e:lamstat})-(\ref{e:lamfirst}) and (\ref{e:rstat})-(\ref{e:psifirst}).
This gives a series of estimates
$E_n^{N,\rm stat}(t,t_0)$, $p_n^{N,\rm stat}(t,t_0)$ etc.
with associated statistical errors,
which we determine by a full Jackknife analysis.

To arrive at final numbers for $E_n$, $p_n$, we first need to estimate
the size of the systematic errors coming from the higher
excited states. To do this, we perform a fit of the form
\be
E_n^{N,\rm stat}(t,t_0) = E_n^{\rm stat} + \beta^{\rm stat}_{n,N} \rme^{-(E^{\rm stat}_{N+1}-E^{\rm stat}_n)t}
\ee
(cf.
\eq{e:lamstatfit})
to the GEVP results for $E_n^{N,\rm stat}(t,t_0)$, fitting the data
at $N=3,\dots,5$, $t_0/a=3,\dots,6$ and $n=1,\dots,6$ simultaneously.
Then, using the values of $E_n^{\rm stat}$ and $\beta^{\rm stat}_{n,N}$
determined from this fit as (fixed) input parameters, we fit
$E_n^{N,\rm kin}(t,t_0)$ and $E_n^{N,\rm spin}(t,t_0)$ by
\bea
E_n^{N,\rm kin}(t,t_0) &=& E_n^{\rm kin} + \left[
\beta^{\rm kin}_{n,N} - \beta^{\rm stat}_{n,N}\,t\,(E^{\rm kin}_{N+1}-E^{\rm kin}_n)
\right]\rme^{-(E^{\rm stat}_{N+1}-E^{\rm stat}_n)t} \\
E_n^{N,\rm spin}(t,t_0) &=& E_n^{\rm spin} + \left[
\beta^{\rm spin}_{n,N} - \beta^{\rm stat}_{n,N}\,t\,(E^{\rm spin}_{N+1}-E^{\rm spin}_n)
\right]\rme^{-(E^{\rm stat}_{N+1}-E^{\rm stat}_n)t}
\eea
(cf.
\eq{e:lamfirstfit})
in the same manner.
Subsequently, we also fit $p_n^{N,\rm stat}(t,t_0)$ to 
\be
p_n^{N,\rm stat}(t,t_0) =  p_n^{\rm stat} + \gamma_{n,N}^{\rm stat}\rme^{-(E^{\rm stat}_{N+1}-E^{\rm stat}_n)t_0}
\ee
(cf.
\eq{e:psistatfit}),
and $p_n^{N,\rm kin}(t,t_0)$, $p_n^{N,\rm spin}(t,t_0)$ and
$p_n^{N,\delta A}(t,t_0)$ to
\bea
p_{n}^{N,\rm kin}(t,t_0) &=& p_n^{\rm kin} +\left[
\gamma_{n,N}^{\rm kin} - \gamma_{n,N}^{\rm stat}p_n^{\rm kin}
-\gamma_{n,N}^{\rm stat}\,t_0\,(E^{\rm kin}_{N+1}-E^{\rm kin}_n)
\right]\rme^{-(E^{\rm stat}_{N+1}-E^{\rm stat}_n)t} \nonumber\\ \\
p_{n}^{N,\rm spin}(t,t_0) &=& p_n^{\rm spin} +\left[
\gamma_{n,N}^{\rm spin} - \gamma_{n,N}^{\rm stat}p_n^{\rm spin}
-\gamma_{n,N}^{\rm stat}\,t_0\,(E^{\rm spin}_{N+1}-E^{\rm spin}_n)
\right]\rme^{-(E^{\rm stat}_{N+1}-E^{\rm stat}_n)t}\nonumber \\ \\
p_n^{N,\delta A}(t,t_0) &=& p_n^{\delta A} +\left[
\gamma_{n,N}^{\delta A} - \gamma_{n,N}^{\rm stat}p_n^{\delta A}
\right]\rme^{-(E^{\rm stat}_{N+1}-E^{\rm stat}_n)t}
\eea
(cf.
\eq{e:psifirstfit}),
using the previously determined values for $E_n^{\rm stat}$,
$\gamma_{n,N}^{\rm stat}$, $E_n^{\rm kin}$ and $E_n^{\rm spin}$
as fixed input parameters.

\begin{figure}
\begin{center}
\includegraphics[height=0.7\textwidth,angle=270]{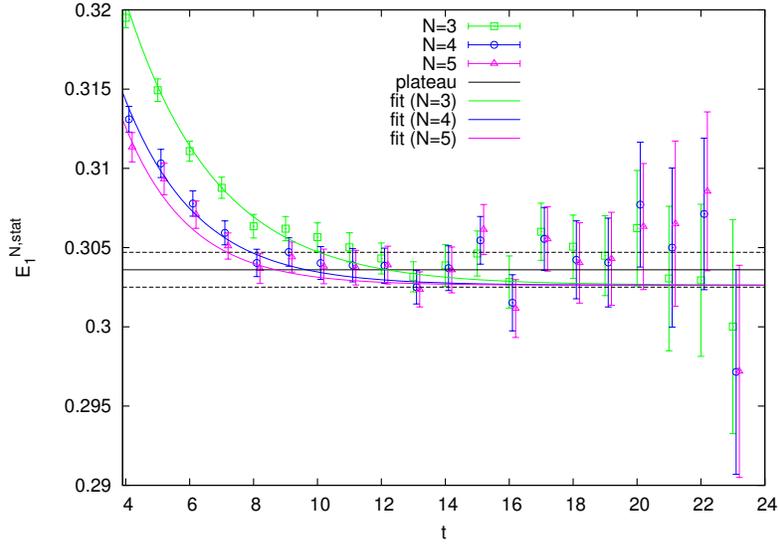}
\end{center}
\caption{Plot of $E_1^{N,\rm stat}(t,t_0=4a)$ against $t$.
         All quantities are given in lattice units.
         Data points are displaced horizontally for better visibility.
         Also shown for the purposes of comparison are
         the fit curves from \eq{e:lamstatfit}
         and the plateau given in table \ref{tab:fitted}.
         Note that the curves are fitted to the total data set,
         not just the points shown in the plot, and that the
         plateau is obtained at a different $t_0$ than shown here.
        }
\label{fig:e1stat}
\end{figure}

\begin{figure}
\begin{center}
\includegraphics[height=0.7\textwidth,angle=270]{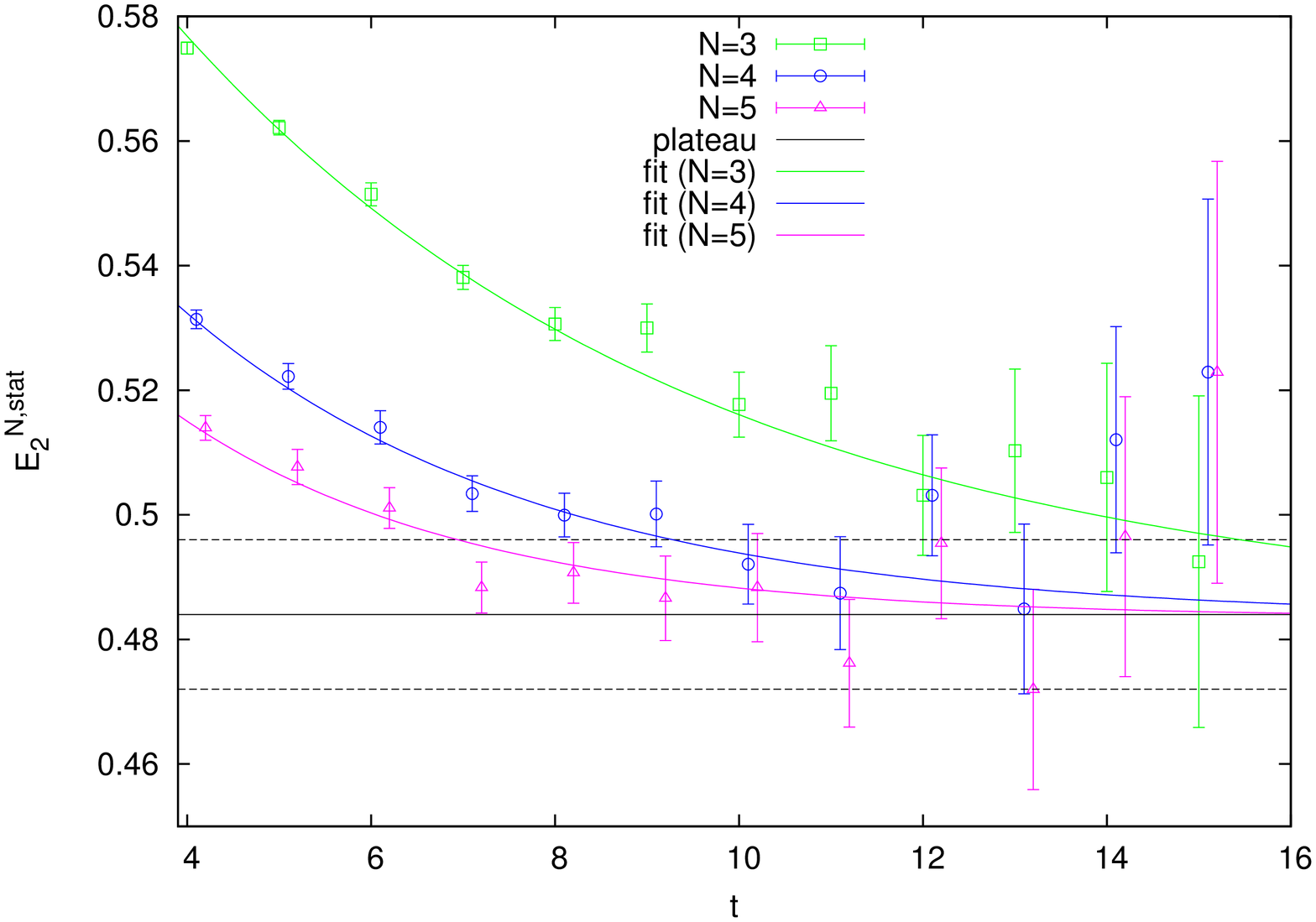}
\end{center}
\caption{Plot of $E_2^{N,\rm stat}(t,t_0=4a)$ against $t$
         in the same style as fig. \ref{fig:e1stat}.
}
\label{fig:e2stat}
\end{figure}

\begin{figure}
\begin{center}
\includegraphics[height=0.7\textwidth,angle=270]{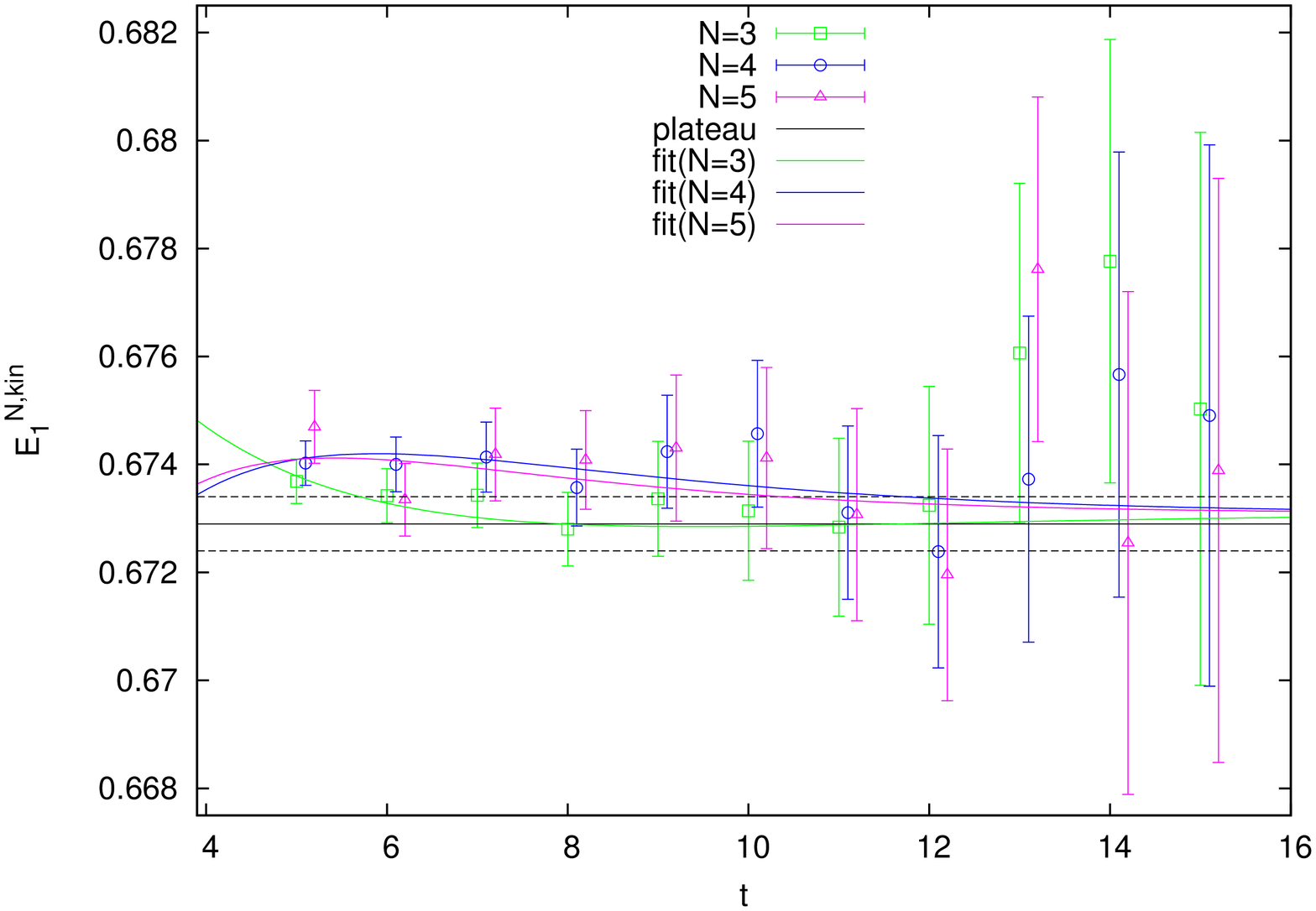}
\end{center}
\caption{Plot of $E_1^{N,\rm kin}(t,t_0=4a)$ against $t$
         in the same style as fig. \ref{fig:e1stat}.
}
\label{fig:e1kin}
\end{figure}

\begin{figure}
\begin{center}
\includegraphics[height=0.7\textwidth,angle=270]{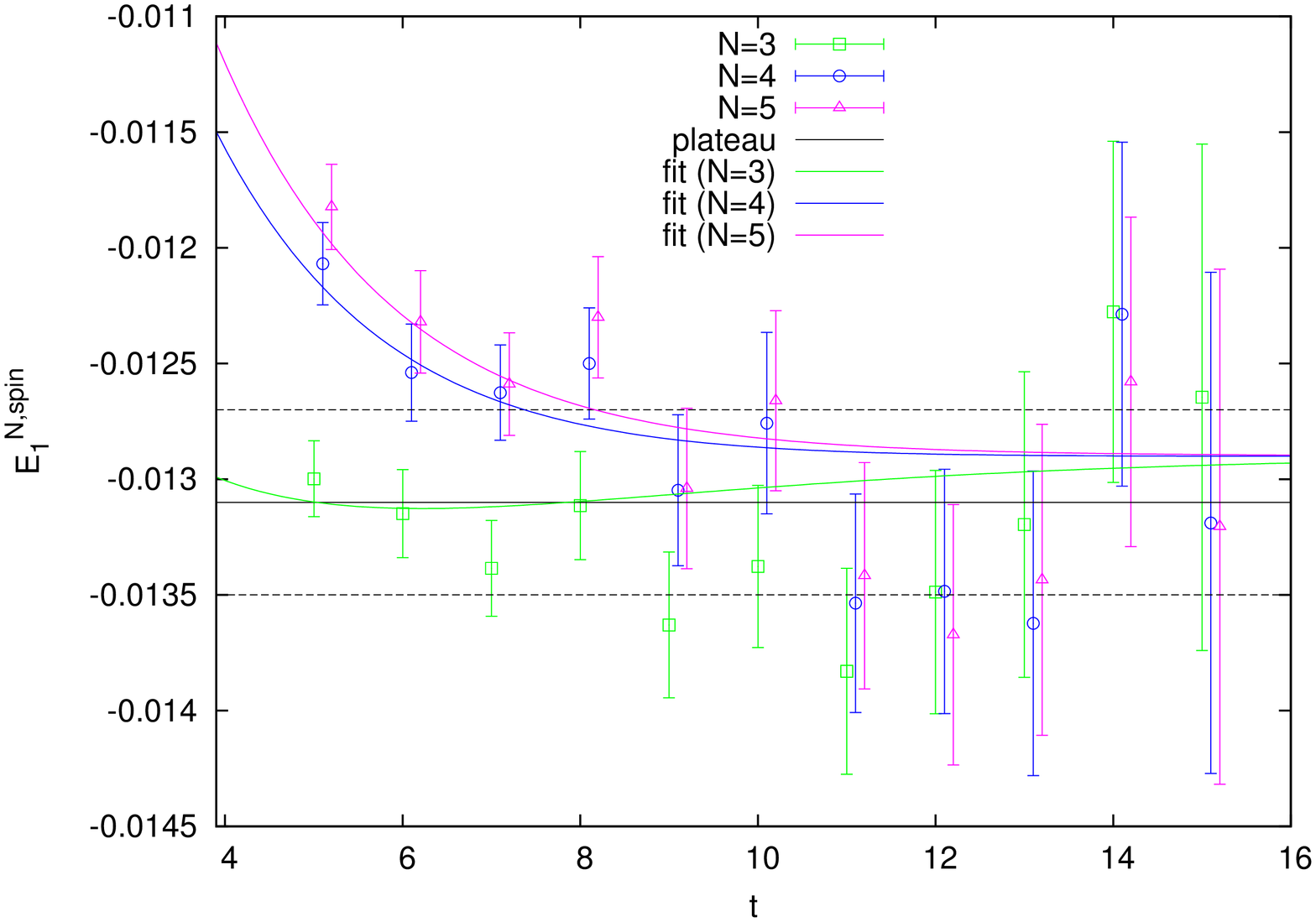}
\end{center}
\caption{Plot of $E_1^{N,\rm spin}(t,t_0=4a)$ against $t$
         in the same style as fig. \ref{fig:e1stat}.
}
\label{fig:e1spin}
\end{figure}

\begin{figure}
\begin{center}
\includegraphics[height=0.7\textwidth,angle=270]{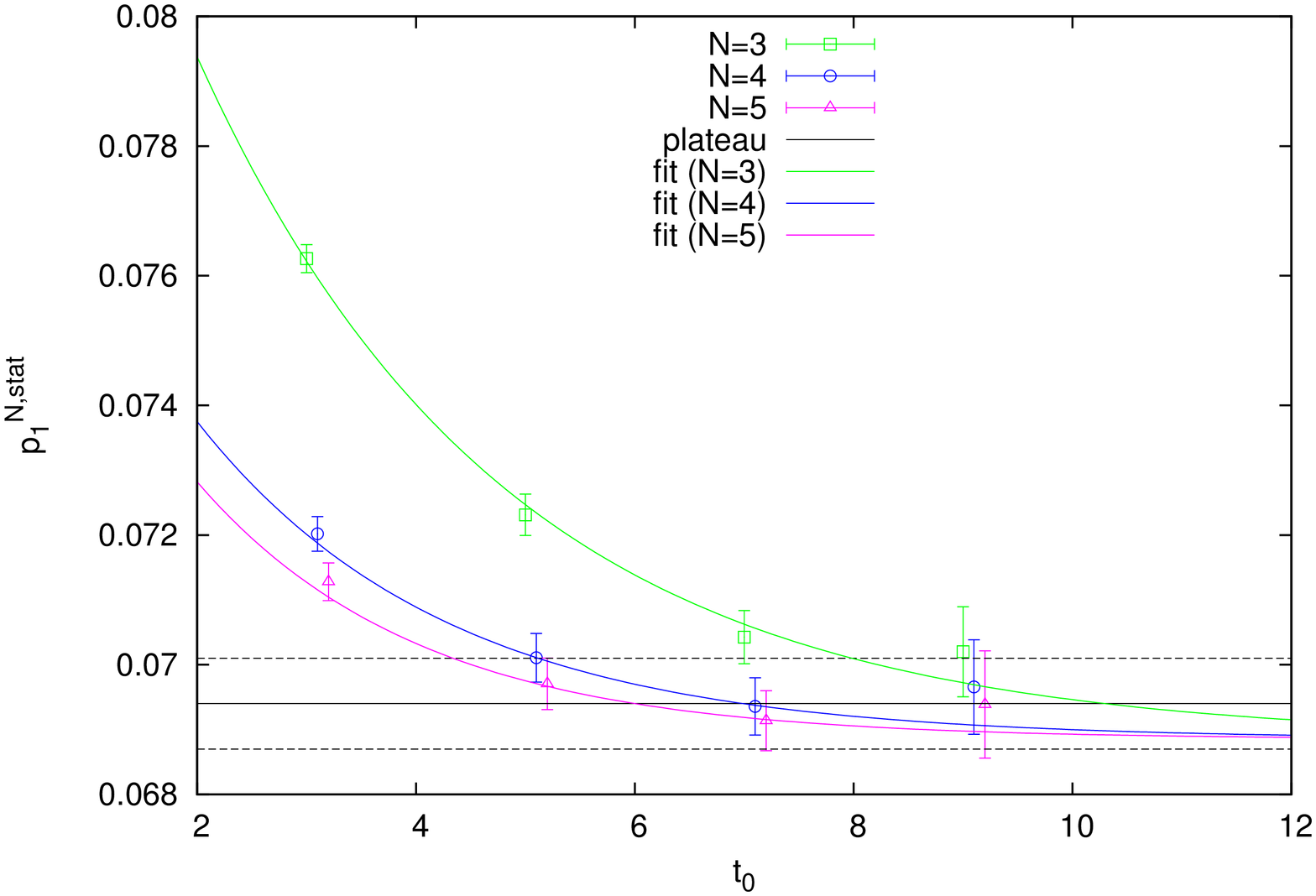}
\end{center}
\caption{Plot of $p_1^{N,\rm stat}(t_0+3a,t_0)$ against $t_0$
         in the same style as fig. \ref{fig:e1stat}.
}
\label{fig:psi1stat}
\end{figure}

\begin{figure}
\begin{center}
\includegraphics[height=0.7\textwidth,angle=270]{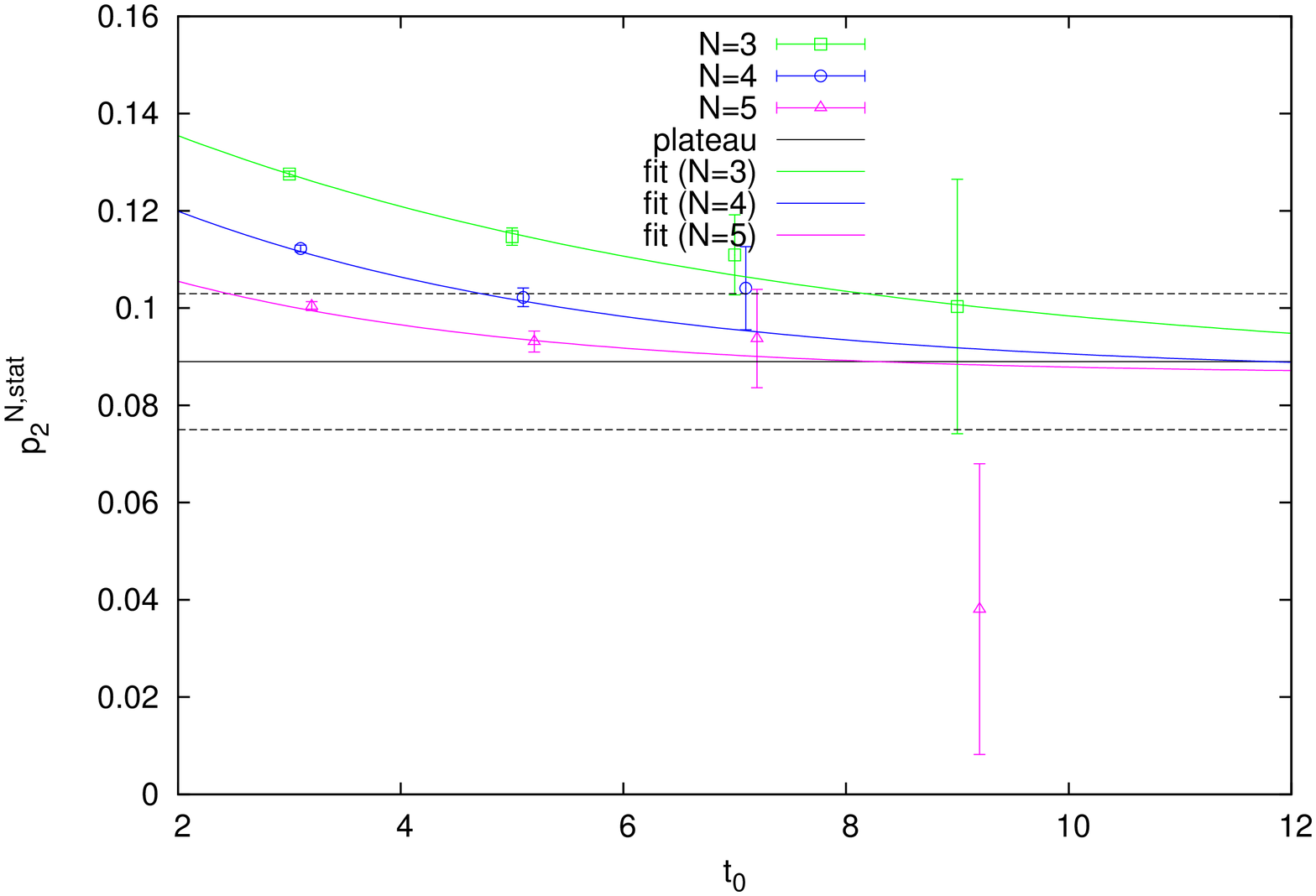}
\end{center}
\caption{Plot of $p_2^{N,\rm stat}(t_0+3a,t_0)$ against $t_0$
         in the same style as fig. \ref{fig:e1stat}.
}
\label{fig:psi2stat}
\end{figure}

\begin{figure}
\begin{center}
\includegraphics[height=0.7\textwidth,angle=270]{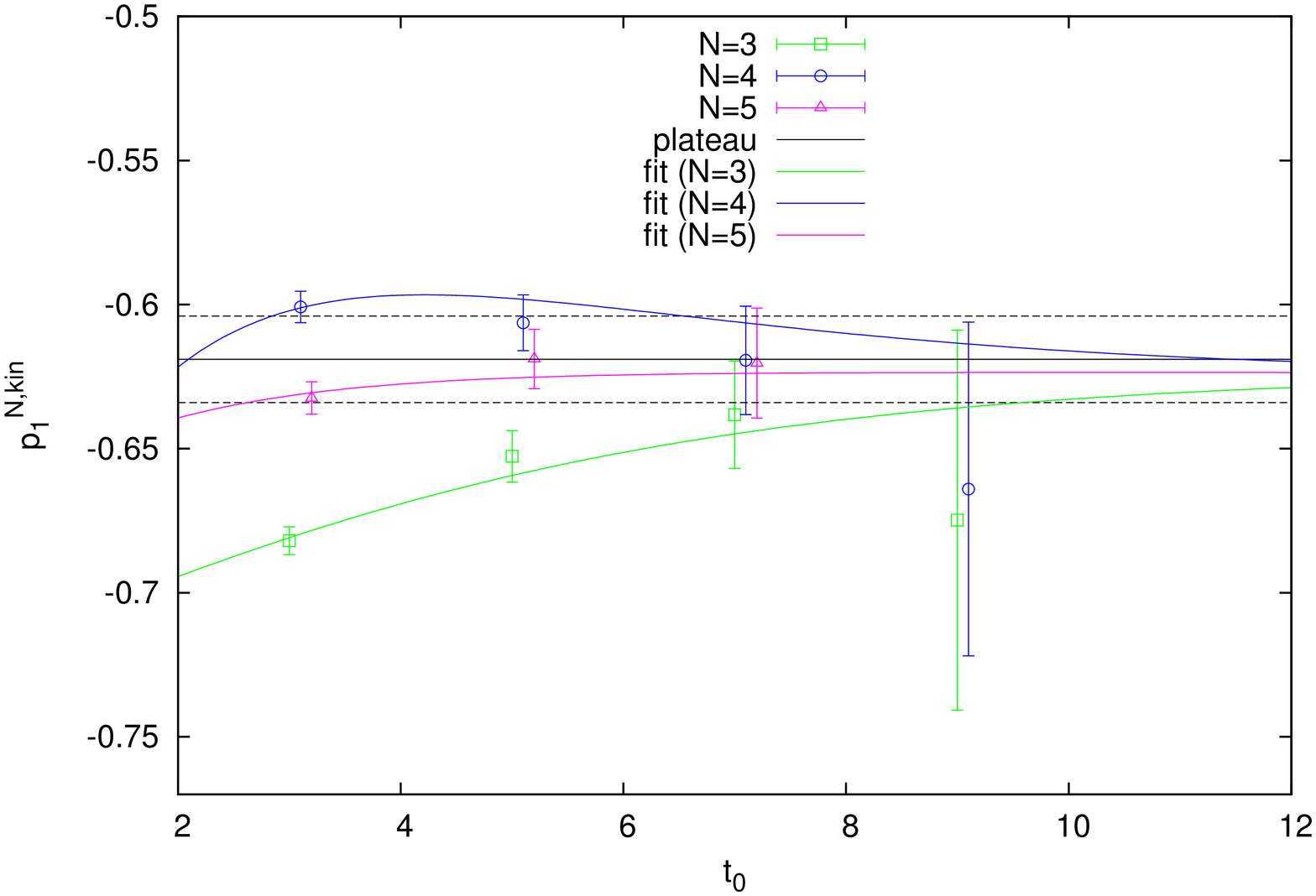}
\end{center}
\caption{Plot of $p_1^{N,\rm kin}(t_0+3a,t_0)$ against $t_0$
         in the same style as fig. \ref{fig:e1stat}.
}
\label{fig:psi1kin}
\end{figure}

\begin{figure}
\begin{center}
\includegraphics[height=0.7\textwidth,angle=270]{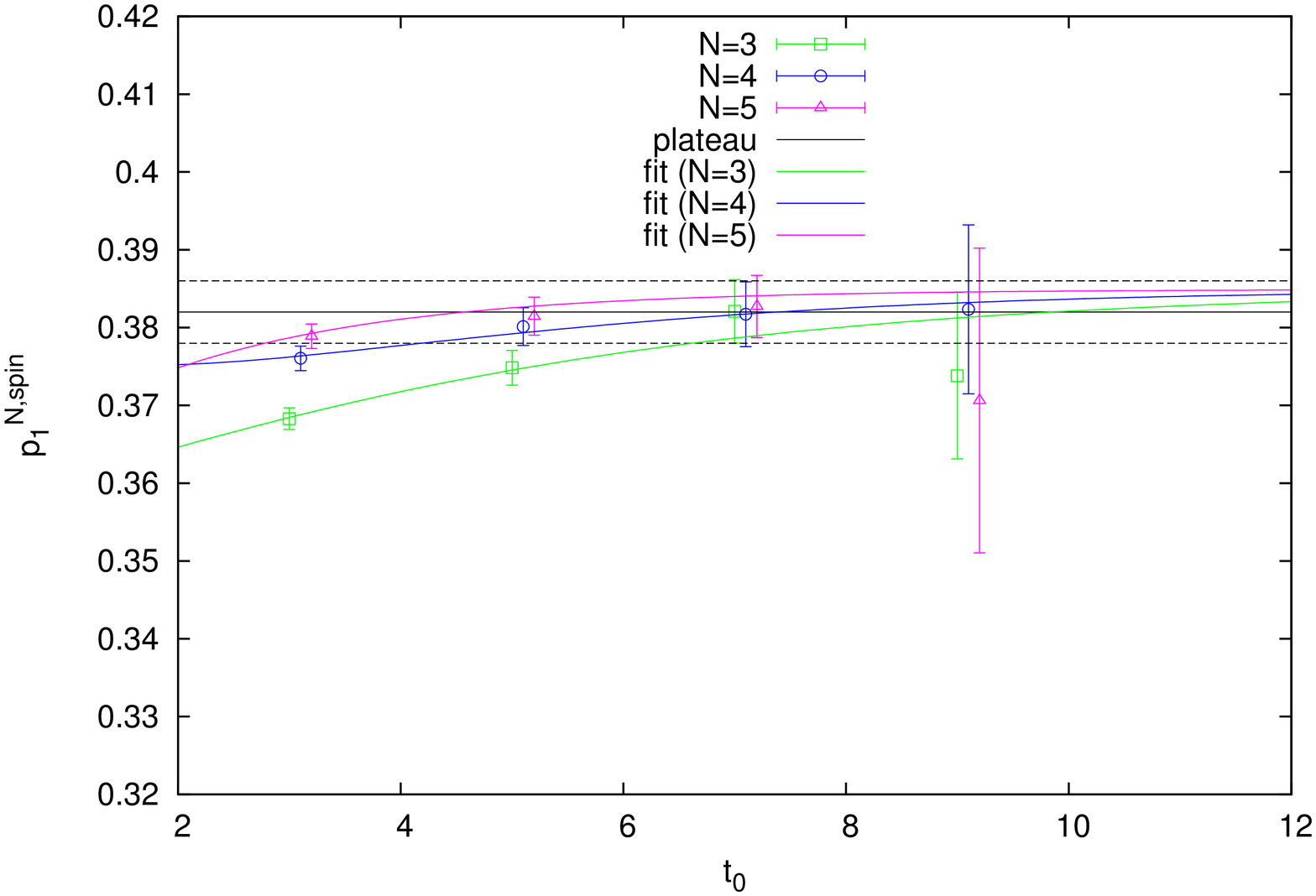}
\end{center}
\caption{Plot of $p_1^{N,\rm spin}(t_0+3a,t_0)$ against $t_0$
         in the same style as fig. \ref{fig:e1stat}.
}
\label{fig:psi1spin}
\end{figure}

Since our fits do not include contributions beyond the leading
excited state correction,
the errors estimated from the fits may not be very reliable.
In particular we consider the fitted values of
$E_4^{\rm stat},\dots,E_6^{\rm stat}$ as rough estimates only,
which may have significant additional systematic errors.
We therefore use the fitted values only for the purpose of determining
the systematic correction $\epsilon(t)$ coming from the excited states.
The good quality of the fit that can be seen from figs.
\ref{fig:e1stat} to \ref{fig:psi1spin}
makes them very suitable for that purpose. In particular we note
that we find the $E_n^N(t,t_0)$ to be essentially independent of
$t_0$, as predicted by our fit formulas.

For a reliable estimate of our quantities of interest,
we calculate plateau averages of the energies
from $t=t_{\rm min}\ge t_0$ to $t=2t_0$ at each $N$ and $t_0$
(and from $t_0=t_{0,\rm min}\ge t/2$ to $t_0=t$ at each
$N$ and $t$ for the matrix elements).
As our final estimate we take that plateau average for which the
absolute sum $\sigma_{\rm tot}=\sigma_{\rm stat}+\sigma_{\rm sys}$
of the statistical error $\sigma_{\rm stat}$ of the plateau average and
the maximum systematic error $\sigma_{\rm sys}=\epsilon(t_{\rm min})$
becomes minimal,
subject to the constraint that $\sigma_{\rm sys}<\sigma_{\rm stat}$.
We impose the latter constraint in order to ensure that the
systematic errors are subdominant.


\begin{table}
\begin{center}
\begin{tabular}{lllll}\hline\hline
Quantity & Full fit & Plateau fit & Plateau & Precision\\\hline
$a E_1^{\rm stat}$ & 0.3026(4) &   0.3036(9)(2) & 5,9,12 & $\ltsim 3$ {\rm MeV}\\ 
$a E_2^{\rm stat}$ & 0.483(3) &   0.484(7)(5) & 5,6,10 & $\sim 20$ {\rm MeV}\\ 
$a E_3^{\rm stat}$ & 0.61(1)  &  0.59(3)(2) & 5,6,11 & $\sim 100$ {\rm MeV}\\
$a E_4^{\rm stat}$ & 0.66(1)  & --- & --- & --- \\
$a E_5^{\rm stat}$ & 0.74(2)  & --- & --- & --- \\
$a E_6^{\rm stat}$ & 0.79(4)  & --- & --- & --- \\
$a^{3/2} p_1^{\rm stat}$ & 0.0689(2) &   0.0694(5)(2) & 5,8,12 & 0.7\%\\ 
$a^{3/2} p_2^{\rm stat}$ & 0.086(3)  &  0.089(10)(4) & 5,7,10 & 12\%\\ 
$a^2 E_1^{\rm kin}$ & 0.6731(4)  &  0.6729(4)(1) & 3,5,6 & $\ltsim 0.6$ {\rm MeV}\\ 
$a^2 E_2^{\rm kin}$ & 0.691(5)  &  0.69(2)(2) & 4,8,11 & $\sim 40$ {\rm MeV}\\ 
$a^2 E_1^{\rm spin}$ & -0.0129(1) &   -0.0131(3)(1) & 4,8,9 & $\ltsim 0.7$ {\rm MeV}\\ 
$a^2 E_2^{\rm spin}$ & -0.0106(4)  &  -0.0113(9)(6) & 5,6,7 & $\sim 2$ {\rm MeV}\\ 
$a p_1^{\rm kin}$ & -0.622(8) &   -0.619(12)(3) & 5,5,8 & 0.6\%\\ 
$a p_1^{\rm spin}$ & 0.385(2)  &  0.382(3)(1) & 5,6,8 & 0.7\%\\
$a p_1^{\delta A}$ & 0.3120(3) & 0.315(3)(1) & 4,6,18 & 0.1\%\\\hline\hline
\end{tabular}
\end{center}
\caption{The values of quantities of interest as determined from
the fit and plateau picking procedure described in the text. The errors
in the column headed ``Plateau fit'' are statistical and systematic.
The column headed ``Plateau'' contains $N$, $t_0$ and $t_{\rm min}$ of
the best plateau for energy levels, and $N$, $t_{0,\rm min}$ and $t$
of the best plateau for matrix elements. The column headed ``Precision''
contains an estimate of the contribution of the total error of that
quantity to the total absolute error on the energy of the state, or to
the total relative error of the decay constant, respectively.
Note that terms such as $\omega_\mrm{kin}\,p_1^\mrm{kin}$ do {\em not}
directly give the physical $1/m$ correction since they contribute divergent 
contributions cancelled for example by $\zahqet$.}
\label{tab:fitted}
\end{table}

Figs.
\ref{fig:e1stat} to \ref{fig:psi1spin}
show plots of our results at different values of $N$ along with the fits
and the optimal plateau and its error bands.
It is apparent not only that the data are very well described
by the fitted functional forms, but also that the fits agree within errors
with the more conservative plateau estimates.

To study how independent the results obtained from the GEVP are from the
operator basis used, we have rerun our analysis for the static case using
a different basis of operators where we exchanged some of the intermediate
smearing levels against operators constructed by an additional application
of the Laplacian. We found that the results for both the static energies
and the static matrix elements were the same within errors as those obtained
from our original basis.

%% file: s4.tex
\section{Discussion}

We have performed a theoretical analysis of the  
GEVP for lattice field theories, \eq{e:gevp}. The $N\times N$ problem
is expanded in terms of a convergent perturbative expansion
around the unperturbed system defined by truncating the
spectral representation of the
correlation matrices after $N$ levels. The contribution 
of all the higher levels defines the perturbation. 

The GEVP
involves two time separations, $t_0,t$, where $t>t_0$. 
At large $t$ and fixed $t_0$, 
the extracted energies $E_n^\mrm{eff}$ converge to the 
eigenvalues of the Hamiltonian with a rate
\bes
  \rmO(\rme^{-\Delta E_n\, t})\,, \quad \label{e:bad}
\ees
where $\Delta E_n$ is the distance of $E_n$ to the closest
energy level. Such a spectral gap is typically
only a few hundred $\MeV$, requiring $t$ to be above 
$t>1\,\fm$ for an appreciable suppression. One realizes,
however, that the first order corrections are only of size
\bes
  \rmO(\rme^{-(E_{N+1}-E_n)\, t})\,, \label{e:good}
\ees
where $E_{N+1}-E_n$ can be much bigger than $\Delta E_n$. 
At the second order in perturbation theory and beyond,
the mixing of different levels introduces the slower
decaying corrections, \eq{e:bad} when $t_0$ is kept fixed,
$t$ is taken large.
We could show that the favorable suppression,  \eq{e:good}, is recovered
to all orders if one chooses $t_0\geq t/2$. This is possible because the 
mentioned mixing
is suppressed exponentially in $t_0$. Thus, even if it is challenging
numerically, a large $t_0$ (and $N$) is more important than a large $t$ to
keep systematic errors small. In the numerical demonstration, section 4, 
we have shown that this property can be used in practice. 
 
Furthermore, because of the exponential suppression of 
mixing with $t_0$, one can also write down interpolating
fields, \eq{e:qeff}, for {\em all} states with energies up to $E_n=E_N$.
These fields are linear combinations of the $N$ fields one started 
from, with coefficients which depend on $t_0$. They are interpolating
fields for the desired energy eigenstates up to corrections
which decay as
\bes
  \rmO(\rme^{-(E_{N+1}-E_n)\, t_0})\,, \label{e:goodmatr}
\ees
again with the large energy gap.

In a numerical example we could demonstrate \eq{e:good} and \eq{e:goodmatr}
nicely by increasing $N,t_0,t$ starting from small values. 
The decrease of the correction
terms is clearly observed, in particular in \fig{fig:e1stat} and \fig{fig:psi1stat}, 
and the fit demonstrates that the derived formulae are (at least
approximately) valid at the accessible values of $N,t_0,t$. This 
quantitative understanding of correction terms allows for a determination
of energy levels and matrix elements with subdominant systematic errors.

In the considered HQET problem we achieved a sub-percent level
(statistical $+$ systematic) determination of the matrix elements needed for
the B-meson decay constant, both at the leading and at the
next-to-leading order in $\minv$. The ground state energy is obtained
at the level of about $3\,\MeV$.
A more complete analysis at several values of $\beta$,
including a continuum extrapolation, is currently in progress.

%% file: a1.tex
\def\azero{A^{(0)}}
\def\aone{A^{(1)}}
\def\bzero{B^{(0)}}
\def\bone{B^{(1)}}
\def\lzero{\lambda^{(0)}}
\def\lone{\lambda^{(1)}}

\section{Recursive perturbative expansion 
\label{a:recursion}}
\subsection{The general case}
Solving \eq{e:pgevp} order by order in $\eps$ and considering the result at
order $\eps^k$, the GEVP reads 
\bes
  0 &=& (\azero-\lzero_n \bzero)v_n^{(k)} 
  \;+\; (\Delta_n -\lone_n \bzero) v_n^{(k-1)} 
  \nonumber\\
  &&+\; (-\lone_n \bone -\lambda^{(2)}_n \bzero) v_n^{(k-2)} 
  \;+\; (-\lambda^{(2)}_n \bone -\lambda^{(3)}_n \bzero) v_n^{(k-3)} 
  \\
  &&+\; \ldots\;
  +\; (-\lambda^{(k-1)}_n \bone -\lambda^{(k)}_n \bzero) v_n^{(0)}\,,
  \nonumber
\ees
where
\be
   \Delta_n = A^{(1)} - \lambda_n^{(0)} B^{(1)}.
\ee
Projecting with $v_n^{(0)}$ gives
\bes
  \label{e:reclam}
  \lambda^{(k)}_n \,\rho_n = (v_n^{(0)}\,,\, \Delta_n v_n^{(k-1)} ) 
  - \sum_{l=1}^{k-1}\lambda^{(l)}_n\, (v_n^{(0)}\,,\, \bone v_n^{(k-1-l)} )\,,
\ees
and a projection with $v_m^{(0)}\,,\;m\neq n$ yields
\bes
  v_n^{(k)} &=& \sum_{m\neq n} \alpha_{mn}^{(k)} v_m^{(0)} 
  \\
  \alpha_{mn}^{(k)} \,\rho_m &=& (v_m^{(0)},\bzero v_n^{(k)}) 
  \nonumber\\ &=&
  {1 \over \lzero_n - \lzero_m} \left\{ (v_m^{(0)},\Delta_n v_n^{(k-1)}) -
          \sum_{l=1}^{k-1}\lambda^{(l)}_n\, (v_m^{(0)}\,,\, \bone v_n^{(k-1-l)} ) 
          \right. 
  \\ && \left. \qquad\qquad\quad \nonumber
          - \sum_{l=1}^{k-1}\lambda^{(l)}_n\, (v_m^{(0)}\,,\, \bzero v_n^{(k-l)} )
          \right\} \,.
\ees
The combined recursions
\be
\label{e:recal}
  \alpha_{mn}^{(k)} \;=\;  \rho_m^{-1}\,  {1 \over \lzero_n - \lzero_m} \left\{ (v_m^{(0)},\Delta_n v_n^{(k-1)}) -
          \sum_{l=1}^{k-1}\lambda^{(l)}_n\,
     \left[ (v_m^{(0)}\,,\, \bone v_n^{(k-1-l)} )
          + \rho_m \alpha_{mn}^{(k-l)}\right] \right\}\,, 
\ee
and \eq{e:reclam} then determine the solution to arbitrary order
in the perturbations starting from the initial values 
$\alpha_{mn}^{(0)}=\delta_{mn}$, $\lambda_{n}^{(0)}(t,t_0)=\rme^{-E_n\,(t-t_0)}$.

\subsection{The case of Euclidean QFT\label{a:algebra}}
We now apply this with 
\bes
  A^{(0)}&=& C^{(0)}(t)\,,\quad \eps A^{(1)}= C^{(1)}(t) \,,\\ 
  B^{(0)}&=& C^{(0)}(t_0)\,,\quad \eps B^{(1)}= C^{(1)}(t_0) \,,\\ 
  v_n^{(0)} &=& u_n\,\quad\to\quad \rho_n=\rme^{-E_{n} t_0}
                                                  \,,
\ees
where we recall that
\be
  C_{ij}^{(0)}(t) \;=\; \sum_{n=1}^N \rme^{-E_n t} \psi_{ni}\psi^*_{nj}\,,
  \quad
  C_{ij}^{(1)}(t) \;=\; \sum_{n=N+1}^\infty \rme^{-E_n t} 
  \psi_{ni}\psi^*_{nj}\,,
\ee
and
\bes
  \psi_{ni} = \langle n| \hat O_i| 0 \rangle\,,\quad \hat H  | n \rangle = E_n | n \rangle\,.
\ees
This means that $\eps v_n^{(1)}, \eps\lambda_n^{(1)}\,\ldots$ are 
corrections to the desired quantities which are
the lowest order $u_n, \lambda_n^{(0)}$. We also recall the orthogonality
\bes
  (u_n,\psi_m) &=& \delta_{mn}\,,\; m,n\leq N\,.
\ees

\subsection{Proof of   eqs.(\ref{e:corren} -- \ref{e:corrp1})  to all orders\label{a:all}}
The first step is to identify what exactly has to be shown concerning the 
behavior of $\lambda_n^{(k)}$ and $\alpha_{mn}^{(k)}$ to prove 
eqs.(\ref{e:corren} -- \ref{e:corrp1}).
For $\corren$ we expand
\bes
  \corren(t,t_0) &=& - \partial_t \log\left(1+\sum_{k\geq1} \eps^k {\lambda_n^{(k)}(t,t_0) 
                                                \over \lambda_n^{(0)}(t,t_0) }\right)
\nonumber \\
          &=& - \sum_{l=1}^{\infty} {(-1)^{l+1} \over l} \sum_{k_1,\ldots,k_l\ge 1} \eps^{\sum_{i}k_i} \partial_t
                \left\{   {\lambda_n^{(k_1)}(t,t_0) \over \lambda_n^{(0)}(t,t_0) }
                          \ldots
                          {\lambda_n^{(k_l)}(t,t_0) \over \lambda_n^{(0)}(t,t_0) }
                \right\}\,. 
\label{e:correnexp}
\ees
If the conditions 
\bes
    \label{e:cond}
    \partial_t {\lambda_n^{(k)}(t,t_0) \over \lambda_n^{(0)}(t,t_0) } &=& 
    \rmO(\rme^{-\Delta E_{N+1,n}\, t}) \,, \quad 
    {\lambda_n^{(k)}(t,t_0) \over \lambda_n^{(0)}(t,t_0) } \;=\; 
    \rmO(1)
\ees
are satisfied,
one sees easily that each term in \eq{e:correnexp} is of order 
$\rme^{-\Delta E_{N+1,n}\, t}$.
In other words \eq{e:cond} is a sufficient condition for \eq{e:corren} to hold.
Next we expand 
\bes
\label{eq:eff_psi1}
  (\aeff)^\dagger |0\rangle = \rme^{-\hat H t} \,(\qeff)^\dagger|0\rangle &=&
\left[ \rme^{E_n t}   \sum_{k\geq0} \epsilon^k\,
       \sum_{n'\geq 1} \rme^{-E_{n'} t} |n'\rangle  (\psi_{n'},v_n^{(k)}(t,t_0))
     \right]\; \times\;\left[1+ \corren \right] \\[2mm]
&& \times \nonumber
\left[1 + \sum_{k_1,k_2\geq1}  \eps^{k1+k2} c_{k_1,k_2}(v_n^{(k_1)}(t,t_0),
                          \rme^{E_n t}C^{(0)}(t) \,v_n^{(k_2)}(t,t_0)) \right.
 \\
        &&\qquad+\, \left.\sum_{k_1,k_2\geq1} \eps^{k1+k2+1}\tilde c_{k_1,k_2}
          (v_n^{(k_1)}(t,t_0), \rme^{E_n t} C^{(1)}(t) \,v_n^{(k_2)}(t,t_0)) \right]
    \nonumber \,,
\ees
with some irrelevant coefficients $c_{k_1,k_2}\,,\;\tilde c_{k_1,k_2}$ and with 
$\corren$ a shorthand for sums of terms $\corren(t',t_0)$, which are all negligible
provided \eq{e:corren} holds.
We express the various terms in \eq{eq:eff_psi1} through $\alpha_{mn}$ 
in order to arrive at conditions for these coefficients:
\bes
  &&\left[ \rme^{E_n t}   \sum_{k\geq0} \,
       \sum_{n'\geq 1} \rme^{-E_{n'} t} |n'\rangle  (\psi_{n'},v_n^{(k)}(t,t_0))
     \right]\nonumber \\ &&\quad=
   \sum_{n'= 1}^N  \sum_{m = 1}^N  \rme^{-(E_{n'}-E_n)t}\,  |n'\rangle\,
               \sum_{k\geq 0}(\psi_{n'},u_m) \alpha_{mn}^{(k)}(t,t_0) 
               + \rmO(\rme^{-\Delta E_{N+1,n}\, t}) \nonumber
  \\ &&\quad=    \sum_{m = 1}^N  \rme^{-(E_{m}-E_n)t}\,  |m\rangle\,
            \sum_{k\geq 0} \alpha_{mn}^{(k)}(t,t_0) 
            + \rmO(\rme^{-\Delta E_{N+1,n}\,t})
  \\ &&\quad= |n\rangle  +\sum_{m \ne n=1}^N  \rme^{-(E_{m}-E_n)t}\, |m\rangle \,
               \sum_{k\geq 1} \alpha_{mn}^{(k)}(t,t_0) + \rmO(\rme^{-\Delta E_{N+1,n}\,t})  \nonumber
\ees
\bes
   (v_n^{(k_1)}, \rme^{E_n t} C^{(1)}(t) \,v_n^{(k_2)}) &=&
    \rmO(\rme^{-\Delta E_{N+1,n}\, t}) 
  \\[1ex]
   (v_n^{(k_1)}, \rme^{E_n t} C^{(0)}(t) \,v_n^{(k_2)}) &=& 0 
   \quad \mbox{if $k_1=0$ or $k_2=0$} 
  \\[1ex]
   (v_n^{(k_1)}, \rme^{E_n t} C^{(0)}(t) \,v_n^{(k_2)}) &=&  
   \sum_{m,m'} \alpha_{mn}^{(k_1)}(t,t_0) \alpha_{m'n}^{(k_2)}(t,t_0) 
    (u_m, \rme^{E_n t} C^{(0)}(t) \,u_m') 
  \nonumber \\  
   &=&  \sum_{m} \alpha_{mn}^{(k_1)}(t,t_0) \alpha_{mn}^{(k_2)}(t,t_0) \rme^{-(E_m-E_n) t}
\ees
A look at these terms shows that it is sufficient to prove 
\bes
   \label{e:cond1}
   \alpha_{mn}^{(k)}(t,t_0) &=& \left\{
            \barr{ll} \rmO(\rme^{-\Delta E_{N+1,m}\, t_0}) \quad &\mbox{for}\quad m>n \\
                  \rmO(\rme^{-\Delta E_{N+1,m}\, t_0} \rme^{-\Delta E_{n,m}\, (t-t_0)}) 
                  \quad &\mbox{for} \quad m<n
            \earr\right.
   \\
   \label{e:cond2}
    \partial_t {\lambda_n^{(k)}(t,t_0) \over \lambda_n^{(0)}(t,t_0) } &=& 
    \rmO(\rme^{-\Delta E_{N+1,n}\, t})  \\
   \label{e:cond3}
    {\lambda_n^{(k)}(t,t_0) \over \lambda_n^{(0)}(t,t_0) } &=& 
    \rmO(1)\,
\ees
for all $k$. (The last two conditions are sufficient to show that 
$\corren(t,t_0)=\rmO(\rme^{-\Delta E_{N+1,n}\, t})$). Note that for $k=0$ the above 
conditions hold trivially.

As a next step we collect the large time behaviour of 
the different terms appearing in the
recursions \eq{e:reclam} and \eq{e:recal} for $\lambda^{(k)}\,,\; \alpha^{(k)}$: 
\bes
  v_n^{(0)} &=& \rmO(1)\,, \quad 
  \\ 
  \eps\bone &=& \rmO(\rme^{-E_{N+1} t_0 })\,, \;
  \eps\aone = \rmO(\rme^{-E_{N+1} t })\,, \\ 
  \eps\Delta_n &=& \rmO(\rme^{-E_{N+1} t_0 }\,
          \rme^{-E_n (t-t_0) })  + \rmO(\rme^{-E_{N+1} t })
          \,,
  \\[2ex]
  \eps\rho_m^{-1}\bone &=& \rmO(\rme^{-(E_{N+1}-E_m) t_0 } )\,, \\ 
  {\eps\rho_m^{-1}\Delta_n\over\lambda_n^{(0)}} &=& \rmO(\rme^{-(E_{N+1}-E_m) t_0 }) 
          + \rmO(\rme^{-(E_{N+1}-E_m) t_0 }\, \rme^{-(E_{N+1}-E_n)(t-t_0) } )\,,
\ees 
where in $\Delta_n$ we keep the two terms because the leading one 
has no $t$-dependence.
For shortness we drop the $\rmO$ symbol from now on 
but just count orders and we 
use $t>t_0$.
This means e.g. that $\rme^{-E_{N+1} t_0 } + \rme^{-E_{n} t_0 } = \rme^{-E_n t_0 }$. 
However, since
derivatives with respect to $t$ are relevant we have to be careful
not to drop $\rme^{-E_{n} t }$ with respect to $\rme^{-E_{n} t_0 }$. 
 
The shorthands
\bes
  \eta_{Nm}(t) &=& \rmO(\rme^{-(E_{N+1}-E_m) t})\,\\
  \gamma_{nm}(t) &=& {\rme^{-E_n t}\over \rme^{-E_n t} - \rme^{-E_m t}}\\
               &=& \left\{ \barr{ll} \sum_{j=0}^\infty \rme^{-j(E_m-E_n) t} \quad &\mbox{when }m>n 
                               \\ -\sum_{j=1}^\infty \rme^{-j(E_n-E_m) t} \quad &\mbox{when }n>m  
                           \earr \right. \,,
\ees
are useful to discuss the large $t,t_0$ asymptotics,
\bes
  \label{e:gammaasympt}
   {\lzero_n \over \lzero_n - \lzero_m} 
   &=& \gamma_{nm}(t-t_0)  
  \;=\; \left\{ \barr{ll} 1 + \rmO(\rme^{-(E_m-E_n) t} )\quad &\mbox{when }m>n 
                       \\  \rmO( \rme^{-(E_n-E_m) t}) \quad &\mbox{when }n>m  
                           \earr \right. \,, 
\\
   {\rho_{m'}^{-1}\,\eps\Delta_n \over \lzero_n - \lzero_m} &=& 
         \eta_{Nm'}(t_0)\,[1+\eta_{Nn}(t-t_0)]\, \gamma_{nm}(t-t_0)
\\
   {\rho_{m'}^{-1}\,\eps\bone \lzero_n \over \lzero_n - \lzero_m} &=& 
         \eta_{Nm'}(t_0)\, \gamma_{nm}(t-t_0)\,.
\ees
In this notation the asymptotics of the first order corrections are
\bes
  {\eps\lambda_n^{(1)}\over \lambda_n^{(0)}} &=& \eta_{Nn}(t_0)\,[1+\eta_{Nn}(t-t_0)]
                 \,,\quad \\
  \eps\alpha_{mn}^{(1)}  &=&  \eta_{Nm}(t_0)\,[1+\eta_{Nn}(t-t_0)]\, \gamma_{nm}(t-t_0)\,,
\ees
while the recursions are 
\bes
  \eps^k\,{\lambda_n^{(k)}\over \lambda_n^{(0)}} &=& 
                    \eps^k\,{\lambda_n^{(1)}\over \lambda_n^{(0)}}\,\alpha^{(k-1)}
            + \eps^{k-1}\,\sum_{l=1}^{k-1}  {\lambda_n^{(l)}\over \lambda_n^{(0)}}
                     \,\eta_{Nn}(t_0)\,\alpha^{(k-1-l)} 
   \nonumber\\
   &=& {\lambda_n^{(1)}\over \lambda_n^{(0)}} \,
       \left\{\eps^{k}\,\alpha^{(k-1)} + 
         \eps^{k-1}\,\sum_{l=1}^{k-1}  {\lambda_n^{(l)}\over \lambda_n^{(1)}}
                     \,\eta_{Nn}(t_0)\,\alpha^{(k-1-l)}
       \right\}\,
\\ 
   \eps^k\,\alpha_{mn}^{(k)}
   &=& \eps^k\,\alpha_{mn}^{(1)}\,\alpha^{(k-1)} + 
          \eps^{k-1}\,\sum_{l=1}^{k-1} {\lambda_n^{(l)}\over \lambda_n^{(0)}}
               \, \eta_{Nm}(t_0)\, \alpha^{(k-1-l)} \, \gamma_{nm}(t-t_0)\,
   \nonumber\\
   && \quad +  \eps^k\,\sum_{l=1}^{k-1} {\lambda_n^{(l)}\over \lambda_n^{(0)}}
               \, \alpha_{mn}^{(k-l)} \, \gamma_{nm}(t-t_0)\,
   \nonumber\\
   &=& \alpha_{mn}^{(1)}\,\left\{ \eps^k\,\alpha^{(k-1)} + 
          \sum_{l=1}^{k-1} {\lambda_n^{(l)}\over \lambda_n^{(1)}}
               \, \left[\eps^{k-1}\,\alpha^{(k-1-l)} \, \eta_{Nn}(t_0)
                +  \eps^k\,\alpha_{mn}^{(k-l)}
              \right]\right\}\,  
\ees
where $\alpha^{(k)} = \max_m \alpha_{mn}^{(k)}$. 
We finally will need the start values of the derivatives
\bes
   \eps\,\partial_t{\lambda_n^{(1)}\over \lambda_n^{(0)}} &=& 
   \eta_{Nn}(t_0)\,\eta_{Nn}(t-t_0)
   = \eta_{Nn}(t)
                 \,,\quad \\
  \eps\,\partial_t\alpha_{mn}^{(1)}  &=&  \eta_{Nm}(t_0)\,
     [\partial_t\gamma_{nm}(t-t_0)\,+\,\partial_t \eta_{Nn}(t-t_0) 
     \, \gamma_{nm}(t-t_0)  \nonumber \\
      && \qquad\qquad +\,  \eta_{Nn}(t-t_0) \, \partial_t \gamma_{nm}(t-t_0)\,]\,
   \nonumber\\
   &=&   \eta_{Nm}(t_0) \,\rme^{-|E_m-E_n|(t-t_0)} \,,
\ees
where   \eq{e:gammaasympt} has been used. 
In particular the relation
\bes
     \eps\,\eta_{Nn}(t_0)\partial_t\alpha_{mn}^{(1)} \leq \eta_{Nn}(t) 
     \quad \mbox{if} \quad t_0 \geq t/2\,
\ees
or equivalently
\bes
     \eps\,\partial_t\alpha_{mn}^{(1)} \leq \eta_{Nn}(t-t_0) 
     \quad \mbox{if} \quad t_0 \geq t/2\,,
\ees
will be relevant.

In terms of the abreviations
\bes
  r_k =  \eps^{k-1}\,{\lambda_n^{(k)}\over \lambda_n^{(1)}}\,,\quad
  x_k =  \eps^k\,\alpha_{mn}^{(k)} \,,\quad
  X_k =  \eps^k\,\alpha^{(k)} = \eps^k\,\max_m \alpha_{mn}^{(k)}
\ees
we want to show that
\bes
   \label{e:conda}
  x_k &=& \eta_{Nm}(t_0)\,\gamma_{nm}(t-t_0)\,, \quad r_k = \rmO(1) \,, \\
   \label{e:condb}
  \partial_t x_k &=& \eta_{Nn}(t-t_0)  \,, \quad 
  \partial_t r_k = \eta_{Nn}(t-t_0)\,,
\ees
which immediately implies
  \eq{e:cond1} to \eq{e:cond3} and thus proves \eq{e:corren} to \eq{e:corrp1}.
For $k=1$, \eq{e:conda} and \eq{e:condb} are satisfied by the start values 
derived above.
The induction steps from $k-1$ to $k$ for $k\geq2$ are very simple 
(remember we assume $t_0 \geq t/2$):
\bes
  r_k &=&  X_{k-1} +  \eta_{Nn}(t_0) \sum_{l=1}^{k-1} r_l X_{k-1-l} 
      \;=\;  \max_m  \eta_{Nm}(t_0)\,\gamma_{nm}(t-t_0) =  \rmO(1) \,,
      \nonumber\\[-1ex] \\
  x_k &=& x_1\,\left\{r_k+  \sum_{l=1}^{k-1} r_l x_{k-l}\right\}\;=\;  \eta_{Nm}(t_0) \,\gamma_{nm}(t-t_0)\,, 
      \\[1ex]
  \partial_t r_k &=&  \partial_t X_{k-1} +  \eta_{Nn}(t_0)
                     \sum_{l=1}^{k-1} \partial_t[r_l\, X_{k-1-l}]
                  \;=\;
                \eta_{Nn}(t-t_0) \,,
      \\[1ex]
  \partial_t x_k &=& \partial_t [x_1\,r_k]+  \sum_{l=1}^{k-1}  \partial_t [x_1\,r_l\, x_{k-l}]  
                \;=\;
                \eta_{Nn}(t-t_0)  \,.
\ees
They conclude our proof.